\pdfoutput=1
\documentclass[PAPER, atlasdraft=false, UKenglish, cernpreprint, texmf, orcidlogo]{atlasdoc}
\usepackage[subcaption, full]{atlaspackage}
\usepackage{atlasbiblatex}

\usepackage{atlasphysics}
\graphicspath{{logos/}{figures/}}

\usepackage{gensymb}
\AtlasTitle{Missing-mass search in forward-proton-tagged dilepton events with the ATLAS detector}

\AtlasAbstract{%
A search is conducted in proton--proton collisions at the Large Hadron Collider for photon-induced production $pp\rightarrow pp+ \gamma \gamma,~(\gamma\gamma\rightarrow VX)$ of a visible particle $V$ decaying into a pair of same-flavour charged leptons ($e^+e^-$ or $\mu^+\mu^-$) and an undetected invisible component $X$. Measurements of the outgoing proton energies by the ATLAS forward proton spectrometer allow the full photon--photon four-momentum to be reconstructed. By subtracting the visible four-momentum of the central system measured with the ATLAS detector, the `missing mass' of any event components not detected in the central region can be reconstructed, enabling the reconstruction of $X$ without knowing its properties, thus allowing the search to be model-independent. A search for a narrow resonance is performed in the missing-mass spectrum between 100~GeV and 900~GeV. The analysis uses data collected in 2017 from proton--proton collisions at a centre-of-mass energy of $\sqrt{s} = 13$~TeV, corresponding to an integrated luminosity of 14.7~fb$^{-1}$. No significant excess over the Standard Model expectation is observed and upper limits at 95\% confidence level are set on the fiducial cross sections for three different signal models in the range between 128 and 2.5~fb. Additionally, model-independent limits are set on the visible cross section of BSM processes, for two sets of selection criteria. Both individual lepton flavour decay channels of the visible boson and a combination of the two channels are considered.
}

\AtlasRefCode{ANA-EXOT-2023-13}

\PreprintIdNumber{CERN-EP-2025-284}

\arXivId{2603.20837}

\HEPDataRecord{169548}

\AtlasJournal{JHEP}
\AtlasJournalRef{JHEP 07 (2026) 236}
\AtlasDOI{10.1007/JHEP07(2026)236}

\AtlasCoverEgroupAnalysisTeam{atlas-exot-2023-13-analysis-team@cern.ch}

\hypersetup{pdftitle={ATLAS document},pdfauthor={The ATLAS Collaboration}}

\begin{document}

\maketitle

\tableofcontents

\section{Introduction}
\label{sec:intro}

The Standard Model (SM) successfully explains most experimental observations in particle physics with high precision, but it is unable to explain some important phenomena, such as the nature of dark matter (DM), the hierarchy of particle masses, and the observed matter--antimatter asymmetry. These motivate searches for phenomena beyond the Standard Model (BSM), prompting extensive programmes at facilities such as the Large Hadron Collider (LHC)~\cite{Evans:2008zzb}. Many dedicated searches are performed to target specific BSM theories, and these are often limited to a restricted phase space which is sensitive to the specific model probed, for example~\cite{EXOT-2023-14, SUSY-2023-10}. In order to probe a wider range of potential models, including new phenomena which may not be described by existing models, generalised searches with much looser restrictions on the final state can be employed. These often use simplified models which make only general assumptions about potential BSM signatures. Model-independent searches can also be performed, which determine visible cross-section limits on potential signals based purely on the observed data and the predicted background.

Proton--proton ($pp$) collisions at the LHC are energetic enough to involve double photon exchange~\cite{PhotonColl1, PhotonColl2, PhotonColl3}, in which the electromagnetic fields around the protons act as coherent sources of photons at low $Q^2$, leading to interactions between photons. These interactions do not change the quantum numbers of the interacting protons, allowing them to remain intact. This produces a signature for exclusive processes in which there are protons in the forward region accompanied by a centrally produced system. The energy lost by these forward protons can be measured with proton spectrometers, enabling a determination of the photon--photon collision energy and giving information about the kinematics of the process. This principle was explored in several feasibility studies~\cite{Beresford:2018pbt,Harland-Lang:2018hmi} and has since been used for SM measurements~\cite{CMS-PPS-17-001,STDM-2018-16,CMS-TOP-21-007,CMS-SMP-21-014} and BSM searches~\cite{CMS-EXO-21-007,EXOT-2019-28}.

In events where the energy loss of each proton is measured, the kinematic information can be exploited to search for undetected particles. This technique relies on the requirement that the protons which emit the centrally interacting photons remain intact and become deflected from the LHC beam due to their energy loss, where they are measured by a proton spectrometer. This allows the reconstruction of the four-momentum of the initial central state. By subtracting from this the four-momentum of the visible final state $V$, measured by the central detector, the four-momentum of any invisible system $X$ can be inferred. This is called the missing-mass method. Since the transverse momenta of the protons and the radiated photons are negligible compared to their longitudinal momenta, the invariant mass, $m_X$, of the undetected particles can be computed as
\begin{align}
\begin{split}
\label{eqn:missing mass calculation}
m_X^2 &= \left(E_{\gamma\gamma} - E_{V}\right)^2 - \left(\overrightarrow{p}_{\gamma\gamma} - \overrightarrow{p}_{V}\right)^2\\
&= \left[\begin{pmatrix} \xi_1 E_p + \xi_2 E_p \\ 0 \\ 0 \\ \xi_1 E_p - \xi_2 E_p \end{pmatrix} - \begin{pmatrix} E_{V} \\ p_x^{V} \\ p_y^{V} \\ p_z^{V} \end{pmatrix}\right]^2,
\end{split}
\end{align}
where $\xi_{1, 2}$ are the fractional energy losses of the two interacting protons, which travel in opposite directions along the LHC beamline, and $E_p$ is the LHC beam energy. Since the missing-mass reconstruction is performed purely with measurements of the other particles involved in the production of $X$, no prior knowledge of the properties or decay channels of $X$ is required, allowing a generic search. The CMS and TOTEM Collaborations applied this method in the process $pp \rightarrow p(\gamma \gamma \rightarrow Z/\gamma + X)p $, where $X$ denotes a generic undetected BSM particle, setting upper limits on the production cross-section~\cite{CMS-EXO-19-009}.

\begin{figure}[ht]
\centering
\includegraphics[width=0.4\linewidth]{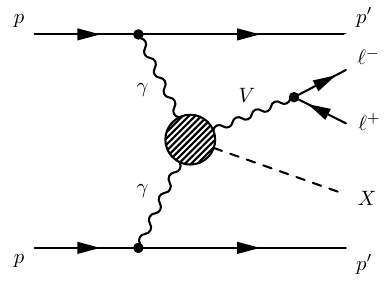}
\caption{Feynman diagram of the elastic signal process, with a leptonically decaying visible boson $V$ produced along with an undetected massive particle or particle system $X$, in association with protons which remain intact, but are deflected away from the main LHC beam.}
\label{fig:missing mass feynman}
\end{figure}

This paper presents a search for missing-mass resonances in $pp$ collision data at a centre-of-mass energy of $\sqrt{s} = 13$~TeV collected in 2017, corresponding to a total integrated luminosity of 14.7~fb$^{-1}$. The analysis uses data from the ATLAS detector, with measurements of forward protons provided by the ATLAS forward proton (AFP) spectrometer~\cite{ATLAS-TDR-24, ATL-PHYS-PUB-2017-012}. The process studied, as illustrated in Figure~\ref{fig:missing mass feynman}, is $pp \rightarrow p(\gamma \gamma \rightarrow \ell\ell + X)p$, where two protons interact exclusively by exchanging photons, which then produce a visible boson decaying leptonically into a pair of electrons or muons, together with an undetected system $X$, which is assumed to decay with a high branching fraction into a system of particles that are invisible to the detector.

The missing mass $m_X$ of the undetected system $X$ is calculated, and the $m_X$ spectrum is searched for a narrow resonance. In this analysis, three different simplified signal models are considered, each characterised by a common final state with two oppositely charged leptons and an undetected system $X$ but different production mechanisms. The first two models involve generic processes in which $X$ is produced along with a $Z$ boson. The third model introduces a specific scenario in which two axion-like particles (ALPs)~\cite{Axion1, ALPLHCSearch} are produced, one short-lived and the other long-lived, corresponding to the $V$ and $X$ systems, respectively. The mass of the $V$ particle is equal to the $Z$ boson mass in all three models, although no specific requirement is placed on the mass of the dilepton system beyond a loose lower limit of 50~GeV. Only the elastic production mode is considered for these signals, as both protons are required to be detected in the AFP spectrometer. Single- and double-dissociative production processes, in which one or both of the signal protons, respectively, break apart into a shower of hadrons following the interaction, are treated as part of the combinatorial background.

This analysis makes use of a track veto in the selection criteria, requiring that no additional tracks with $p_\textrm{T} > 500$~MeV in the central detector are associated with the dilepton vertex. This criterion markedly reduces quark-induced backgrounds, which are dominant in the analysis, allowing an improvement in sensitivity over the previous CMS analysis~\cite{CMS-EXO-19-009} for a common benchmark signal, in the low mass range.
\section{The ATLAS detector and the ATLAS forward proton spectrometer}
\label{sec:atlas_afp}
%

%
\newcommand{\AtlasCoordFootnote}{%
ATLAS uses a right-handed coordinate system with its origin at the nominal interaction point (IP)
in the centre of the detector and the \(z\)-axis along the beam pipe.
The \(x\)-axis points from the IP to the centre of the LHC ring,
and the \(y\)-axis points upwards.
Polar coordinates \((r,\phi)\) are used in the transverse plane,
\(\phi\) being the azimuthal angle around the \(z\)-axis.
The pseudorapidity is defined in terms of the polar angle \(\theta\) as \(\eta = -\ln \tan(\theta/2)\) and is equal to the rapidity
$ y = (1/2) \ln[(E + p_z)/(E - p_z)] $ in the relativistic limit.
Angular distance is measured in units of \(\Delta R \equiv \sqrt{(\Delta y)^{2} + (\Delta\phi)^{2}}\).}

The ATLAS detector~\cite{PERF-2007-01} at the LHC covers nearly the entire solid angle around the collision point.\footnote{\AtlasCoordFootnote}
It consists of an inner tracking detector surrounded by a thin superconducting solenoid, electromagnetic and hadronic calorimeters,
and a muon spectrometer incorporating three large superconducting air-core toroidal magnets.

The inner-detector system (ID) is immersed in a \qty{2}{\tesla} axial magnetic field
and provides charged-particle tracking in the range \(|\eta| < 2.5\).
The high-granularity silicon pixel detector covers the vertex region and typically provides four measurements per track,
the first hit generally being in the insertable B-layer installed before Run~2~\cite{ATLAS-TDR-19,PIX-2018-001}.
It is followed by the silicon microstrip detector, which usually provides eight measurements per track.
These silicon detectors are complemented by the transition radiation tracker (TRT),
which enables radially extended track reconstruction up to \(|\eta| = 2.0\).
The TRT also provides electron identification information
based on the fraction of hits (typically 30 in total) above a higher energy-deposit threshold corresponding to transition radiation.

The calorimeter system covers the pseudorapidity range \(|\eta| < 4.9\).
Within the region \(|\eta|< 3.2\), electromagnetic calorimetry is provided by barrel and
endcap high-granularity lead/liquid-argon (LAr) calorimeters,
with an additional thin LAr presampler covering \(|\eta| < 1.8\)
to correct for energy loss in material upstream of the calorimeters.
Hadronic calorimetry is provided by the steel/scintillator-tile calorimeter,
segmented into three barrel structures within \(|\eta| < 1.7\), and two copper/LAr hadronic endcap calorimeters.
The solid angle coverage is completed with forward copper/LAr and tungsten/LAr calorimeter modules
optimised for electromagnetic and hadronic energy measurements, respectively.

The muon spectrometer comprises separate trigger and
high-precision tracking chambers measuring the deflection of muons in a magnetic field generated by the superconducting air-core toroidal magnets.
The field integral of the toroids ranges between \num{2.0} and \qty{6.0}{\tesla\metre}
across most of the detector.
Three layers of precision chambers, each consisting of layers of monitored drift tubes, cover the region \(|\eta| < 2.7\),
complemented by cathode-strip chambers in the forward region, where the background is highest.
The muon trigger system covers the range \(|\eta| < 2.4\) with resistive-plate chambers in the barrel, and thin-gap chambers in the endcap regions.

The luminosity is measured mainly by the LUCID-2~\cite{LUCID2} detector, which records Cherenkov light produced in the quartz windows of photomultipliers located close to the beam pipe.

Events were selected by the first-level trigger system implemented in custom hardware,
followed by selections made by algorithms implemented in software in the high-level trigger~\cite{TRIG-2016-01}.
The first-level trigger accepted events from the \qty{40}{\MHz} bunch crossings at a rate close to \qty{100}{\kHz},
which the high-level trigger reduced in order to record complete events to disk at about \qty{1.25}{\kHz}.

A software suite~\cite{SOFT-2022-02} is used in data simulation, in the reconstruction
and analysis of real and simulated data, in detector operations, and in the trigger and data acquisition
systems of the experiment.

%
%
%
%

%
%
%
%

\begin{figure}[ht]
\centering
\includegraphics[width=0.7\linewidth]{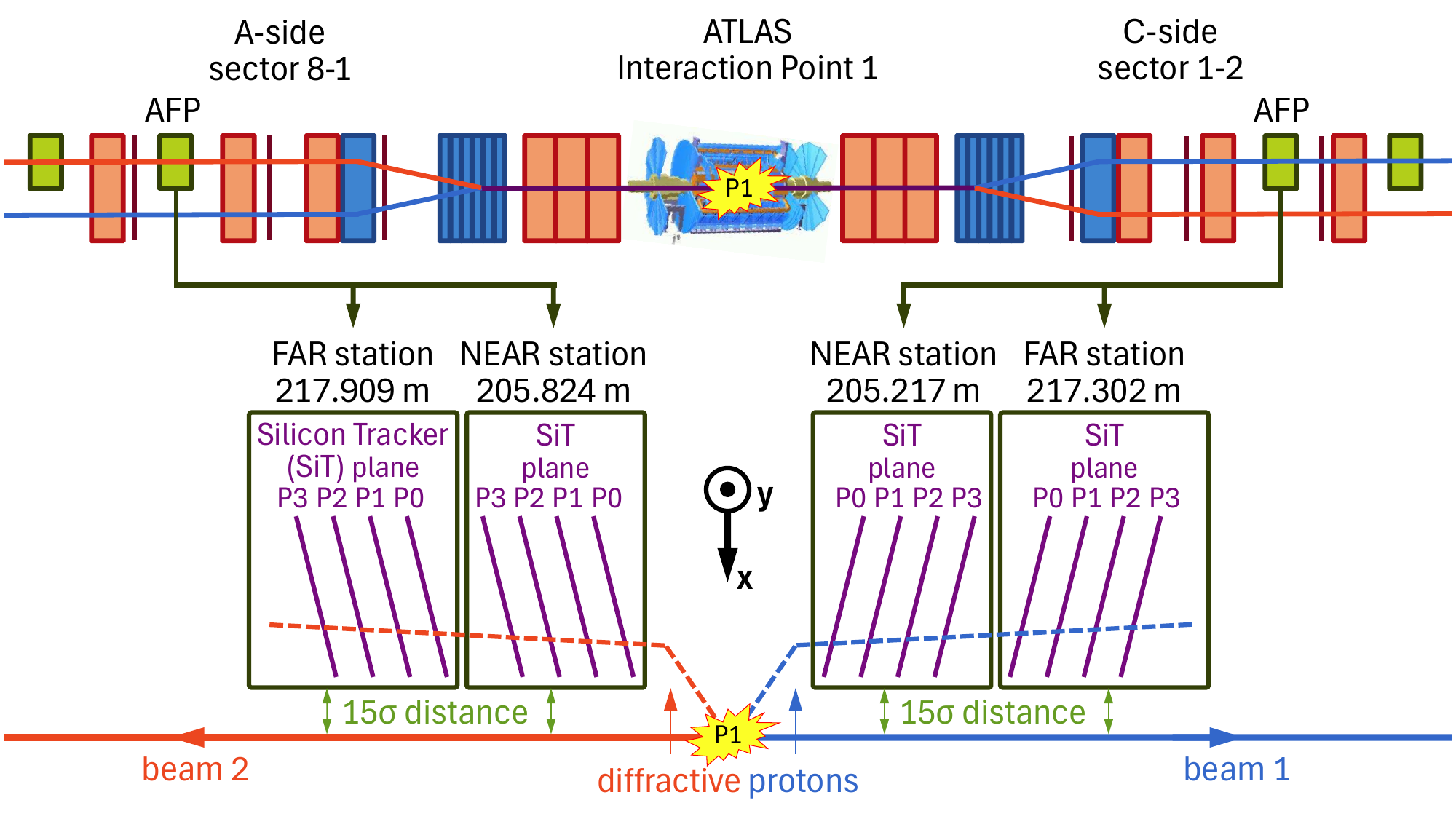}
\caption{Layout of the ATLAS forward proton spectrometer used to detect protons which remain intact following interactions in the central ATLAS detector and become deflected from the main LHC beam. The silicon tracker modules are brought to about $15\sigma$ from the beam centre, where $\sigma$ is the width of the Gaussian beam profile. The orange and blue structures represent quadrupole focusing and dipole bending magnets, respectively.}
\label{fig:afp layout}
\end{figure}

Protons that lose energy but remain intact after colliding at the ATLAS interaction point (IP) are deflected from the LHC beam by the magnetic fields from the beam steering magnets. They are detected using the AFP spectrometer, shown in Figure~\ref{fig:afp layout}. The AFP spectrometer is an ATLAS subdetector consisting of four tracking stations, two on either side of the ATLAS IP, located at longitudinal positions $z = \pm 205$~m and $z = \pm 217$~m from the interaction point, called the \textsc{Near} and \textsc{Far} stations respectively.
Each station contains a silicon tracker (SiT), designed to measure the displacement of deflected protons from the beamline, thereby determining their energy loss in the central interaction. The tracking detectors consist of four planes of edgeless silicon pixel sensors with $336 \times 80$ pixels, where each pixel measures $50 \times 250$~$\mu$m$^2$~\cite{AFP_SiT, AFP_BeamTests, AFP_SiTReadout, AFP_SiTReadout2}. To enhance the hit reconstruction efficiency, the planes are tilted by $14^\circ$ around the $x$-axis, increasing the probability that an incident proton produces hits in at least two planes and resulting in spatial resolutions of $\sigma_x = 6$~$\mu$m and $\sigma_y = 30$~$\mu$m~\cite{AFP_ModuleProd}. Each station is mounted on a movable \enquote{Roman pot} (RP)~\cite{RP1, RP2}, allowing the tracker to be moved in the $x$ direction towards the beam for data-taking, and safely removed at other times to protect the modules from radiation damage. During operation with stable beams, the trackers are positioned approximately 2~mm from the beam centre, corresponding to about $15\sigma$, where $\sigma$ is the Gaussian width of the LHC beam profile. Global alignment of the AFP stations with the central ATLAS detector is performed using beam loss monitors~\cite{AFP_BBA, LHC_BLM}, beam position monitors~\cite{LHC_BPM}, and further in situ calibration using studies of exclusive dimuon production events~\cite{ATL-FWD-PUB-2024-001}.

The signals from each of the struck SiT planes are grouped into clusters and fitted to construct tracks. Small corrections of the order of $0.1$~mm are applied to the cluster coordinates to ensure that cluster positions in different planes are consistent given the spatial resolution. The proton displacement and slope in the $x$-axis are determined from the spatial track coordinates, which are then translated into a measurement of the fractional energy loss $\xi$ of the proton using a parameterisation obtained from LHC optics simulations. The $\xi$ parameter is related to the energy of the deflected proton $E_{\text{p'}}$ as
\begin{equation*}
\label{eqn:proton xi}
\xi = 1 - \frac{E_{\text{p'}}}{E_{\text{beam}}},
\end{equation*}
where $E_{\text{beam}}$ is the nominal beam energy, 6.5~TeV during \RunTwo. By default, protons are reconstructed only from matched tracks in both the \textsc{Near} and \textsc{Far} stations on a given AFP side, called \enquote{double-sided} reconstruction. However, the size of the proton sample can be increased by \enquote{single-sided} reconstruction, using only a \textsc{Far} station track. The AFP acceptance in $\xi$ depends on both the horizontal coverage of the silicon tracker and the LHC beam optics. The lower bound on $\xi$ is determined by how close the silicon tracker sensors come to the beam, while the upper bound is set by the beam collimators located between the interaction point and the RPs. Reconstructed protons in the final signal region are required to have $0.035 < \xi < 0.08$ to match the region where the detector efficiency is well understood. In this region, the proton reconstruction efficiency is measured to be $(92\pm2)\%$, with a $\xi$ resolution of 10\%~\cite{ATL-FWD-PUB-2024-001}. The $\xi$ resolution is measured by comparing central and forward measurements, and therefore includes any effect of the 15~GeV uncertainty in the initial LHC beam energy.

\section{Data and simulated event samples}
\label{sec:data_mc}

\subsection{Data}
The dataset used in this analysis was collected in 2017 from $pp$ collisions at a centre-of-mass energy of $\sqrt{s} = 13$~TeV. In addition to the standard ATLAS data-quality requirements~\cite{DAPR-2018-01}, every AFP station was required to be operational, and the AFP data acquisition system~\cite{AFP_Trigger} had to not report any problems. The resulting total integrated luminosity is $14.7$~fb$^{-1}$. Events were required to pass either a single-lepton or dilepton trigger, applying the lowest available $p_\textrm{T}$ thresholds.  The single-lepton triggers accepted muons or electrons with $p_\textrm{T} > 26$~GeV. The dielectron trigger accepted electrons with $p_\textrm{T} > 17$~GeV, while the dimuon trigger accepted muons with $p_\textrm{T} > 14$~GeV~\cite{ATL-DAQ-PUB-2018-002, TRIG-2018-05, TRIG-2018-01}.

\subsection{Signal models}
\label{subsec:signal models}
Signal events were simulated using different Monte Carlo (MC) generators, depending on the theoretical model. Three different signal models are considered, as described in this subsection. In each case, the response of the central ATLAS detector to the simulated signal was modelled using a full detector simulation~\cite{SOFT-2010-01} based on \GEANT~\cite{Agostinelli:2002hh}. The response of the AFP spectrometer was modelled using a fast simulation in which Gaussian smearing matched to the AFP spatial resolution is applied to the track positions, and the 92\% proton reconstruction efficiency measured in 2017 is included. The effect of multiple interactions in the same and neighbouring bunch crossings (\pileup) was modelled for all signal samples by overlaying the simulated event with inelastic proton--proton (\(pp\)) events generated with \PYTHIA[8.186]~\cite{Sjostrand:2007gs} using the \NNPDF[2.3lo] set of parton distribution functions (PDF)~\cite{Ball:2012cx} and the A3 set of tuned parameters (tune)~\cite{ATL-PHYS-PUB-2016-017}. Events were weighted to reproduce the distribution of the average number of interactions per bunch crossing (\(\left<\mu \right>\)) observed in data.

\begin{figure}[ht]
\centering
\begin{subfigure}{.45\textwidth}
\includegraphics[width=0.95\linewidth]{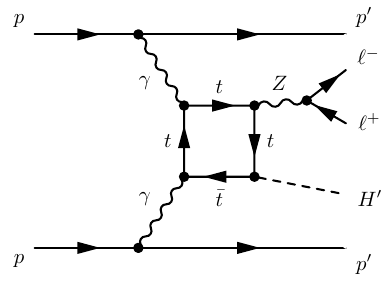}
\end{subfigure}
\begin{subfigure}{.45\textwidth}
\includegraphics[width=0.95\linewidth]{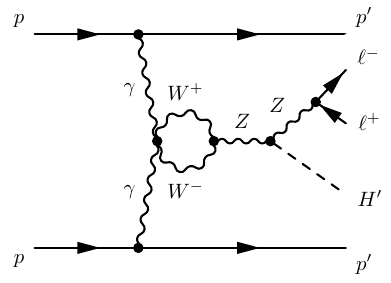}
\end{subfigure}
\caption{Representative Feynman loop diagrams contributing to photon-induced $Z+H'$ production.}
\label{fig:diagram_zh}
\end{figure}

\paragraph{$Z+H'$ model:}
This generic model is based on the SM process $\gamma \gamma \rightarrow ZH$. The Higgs boson is replaced by a generic scalar denoted by $H'$, which is assigned a tunable mass that matches the missing mass under investigation and is forced to not decay, thus emulating an invisible final-state particle. The $Z$ boson is forced to decay only to leptons ($e^+e^-$, $\mu^+\mu^-$ or $\tau^+\tau^-$) with all three flavours simulated in the same sample, and all couplings are kept at their default SM values for simplicity. The sample generated at a signal mass of 400~GeV exceptionally includes all $Z$ boson decay channels in the SM (e.g. including hadronic decays), for investigations during the analysis. This process can occur via loop-induced interactions, two of which are illustrated in Figure~\ref{fig:diagram_zh}.
Several samples were produced, varying the missing mass $m_X$ between 100 and 800~GeV in 100~GeV steps. Event generation was performed in \MGNLO[2.9.5]~\cite{Alwall:2014hca} using the \texttt{loop\_qcd\_qed\_sm} model \cite{MG_ZH_model}, with the CT14qed\_proton PDF set~\cite{CT14QED}, interfaced to \Pythia[8.306]~\cite{Bierlich:2022pfr} for modelling of the parton showering (PS)~\cite{Mrenna:2016sih}, hadronisation and underlying event. The A14 tune~\cite{ATL-PHYS-PUB-2014-021} of \PYTHIA[8] was used with the \NNPDF[2.3QED] PDF set~\cite{NNPDF23QED}.

\begin{figure}[ht]
\centering
\includegraphics[width=0.44\linewidth]{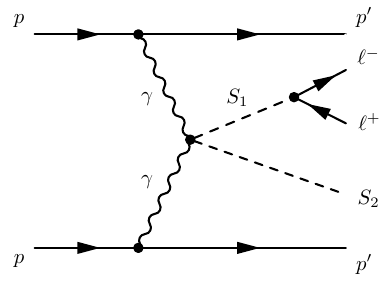}
\caption{Feynman diagram for photon-induced di-ALP ($S_1, S_2$) production.}
\label{fig:ALP_feyn}
\end{figure}

\paragraph{Di-ALP model:}
In this model, two distinct, electrically neutral, scalar ALPs are produced via a photon-induced process~\cite{feynrules,DiALPInspiration}. ALPs are hypothetical scalar or pseudoscalar particles, extensions of the original pseudoscalar axion proposal, and can be viable dark-matter candidates~\cite{ALPLHCSearch, Axion1, AxionDM1}. A short-lived ALP ($S_1$), which is leptophilic and decays into either an $e^+e^-$ or $\mu^+\mu^-$ pair with equal probability ($e^+e^-$ and $\mu^+\mu^-$ events are simulated in the same sample for each signal mass), is produced in association with a long-lived ALP ($S_2$), which does not decay in the detector and escapes detection. The ALPs couple to photons via dimension-five operators containing derivatives~\cite{ALPLHCSearch}. The final state consists of two oppositely charged leptons from $S_1$ and missing mass from the undetected $S_2$, closely mimicking the signature of the generic models. The Feynman diagram for this process is shown in Figure~\ref{fig:ALP_feyn}.
The ALPs $S_1$ and $S_2$ are neutral pseudoscalars, such that $S_2$ behaves similarly to $H'$ described in the previous model, while $S_1$ differs from the $Z$ boson in terms of spin and potential decay modes. Besides this difference, the final state is produced via a four-point interaction, similarly to the $Z+X$ process. The mass of the short-lived ALP is arbitrarily set to that of the Z boson in order to simplify the analysis and allow common selections for all signal models. Several tests were performed with short-lived ALP masses in the range 10--200~GeV without kinematic selections applied and minimal impact on the final kinematic distributions was observed. The mass of the long-lived ALP is scanned from 200~GeV to 800~GeV. Event generation was performed in \MGNLO[2.9.5] using the \texttt{yyALP\_UFO} model \cite{MG_ALP_model}, with the CT14qed\_proton PDF set, interfaced to \Pythia[8.306] for modelling of the parton showering (PS), hadronisation and underlying event. The A14 tune of \PYTHIA[8] was used with the \NNPDF[2.3QED] PDF set.

\begin{figure}[ht]
\centering
\includegraphics[width=0.45\linewidth]{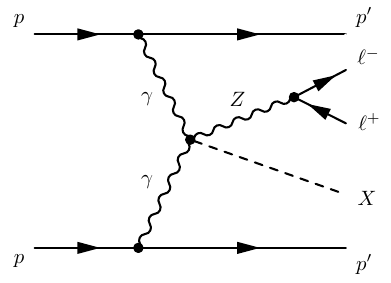}
\caption{Feynman diagram for photon-induced $Z+X$ via a four-point interaction.}
\label{fig:diagram_sc}
\end{figure}

\paragraph{$Z+X$ model:}
This simplified model, proposed in the previous CMS analysis~\cite{CMS-EXO-19-009}, describes the photon-induced production of a $Z$ boson in association with an invisible particle $X$, via a four-point interaction, as illustrated in Figure~\ref{fig:diagram_sc}. The $Z$ boson decays leptonically into electrons or muons, and separate samples are generated with decays to each individual lepton flavour. The $ZX$ invariant mass was generated with a probability proportional to $\textrm{e}^{-\tau \cdot m_{ZX}}$, where $\tau$ is a model parameter fixed at 0.04~GeV$^{-1}$. This $\tau$ value was chosen for consistency with the parameter used in the previous CMS analysis~\cite{CMS-EXO-19-009}, enabling access to a phase space that is not accessible with the other models considered in this analysis. Lowering the value of $\tau$ yields a spectrum similar to those from the other two models, while increasing $\tau$ results in a softer spectrum, thus reducing the selection efficiency. The mass of the new resonance, $m_X$, was varied between 300 and 900~GeV, the range in which the reconstruction efficiency is maximised. The key difference between this model and the $Z+H'$ scenario discussed earlier is the production mechanism: while the $Z+H'$ final state is generated via a loop process, the $Z+X$ final state arises from a direct four-point interaction. Events were generated at the parton level using \textsc{SuperChic}~5.1~\cite{SuperChic}, using a structure function approach~\cite{StructureFunction}. The models used are Processes 83 ($Z\rightarrow\mu\mu$) and 84 ($Z\rightarrow ee$) in the \textsc{SuperChic} manual \cite{SC_Manual}. Modelling of the parton showering, hadronisation and underlying event is performed by \Pythia[8.310]. Since this is a simplified model, no other parameters such as the spin or helicity of the final-state particles are explicitly defined. Therefore, both the $Z$ and $X$ are emitted isotropically in the diphoton centre of mass frame, with random polarisation.

In exclusive production processes, the probability that no additional particles are produced in soft proton--proton interactions is characterised by the soft-survival factor. This factor accounts for the probability that additional proton--proton interactions may occur following the photon exchange, filling with additional activity the low- to mid-rapidity region, which is otherwise empty in exclusive processes~\cite{SoftSurvivalFactor}. This probability has a value $S \leqslant 1$ which depends on the event kinematics. While this effect is implemented in the \textsc{SuperChic} generator for some processes, the simplified $Z+X$ model does not include soft-survival effects at the generation level. Instead, a correction derived from exclusive $\gamma\gamma \rightarrow \ell\ell$ events generated in \textsc{SuperChic} is used to estimate the survival probability $S$ as a function of the mass of the final-state particle system. The obtained relation is well described by a linear parameterisation $S = 0.9381 - 0.000356 \cdot m_{VX}$, corresponding to applied soft-survival factors which vary from 0.80 to 0.58 across the mass range studied. These probabilities are then used to weight the $Z+X$ signal sample event-by-event as a function of the generator-level $m_{ZX}$, and thus obtain the nominal $Z+X$ signal prediction used throughout the analysis. Unlike \textsc{SuperChic}, \textsc{MadGraph} does not provide a built-in mechanism to model soft-survival probabilities. Consequently, no survival correction is applied to either of models generated with \textsc{MadGraph}, resulting in an overestimation of the exclusive signal yield. To account for this limitation and avoid underestimating the theoretical uncertainty, a dedicated systematic uncertainty is assigned to the signal normalisation, reflecting the possible impact of soft interactions on the signal yield. This is discussed further in Section~\ref{sec:sys_unct}.

\begin{figure}[ht]
\centering
\begin{subfigure}{.45\textwidth}
\includegraphics[width=0.95\linewidth]{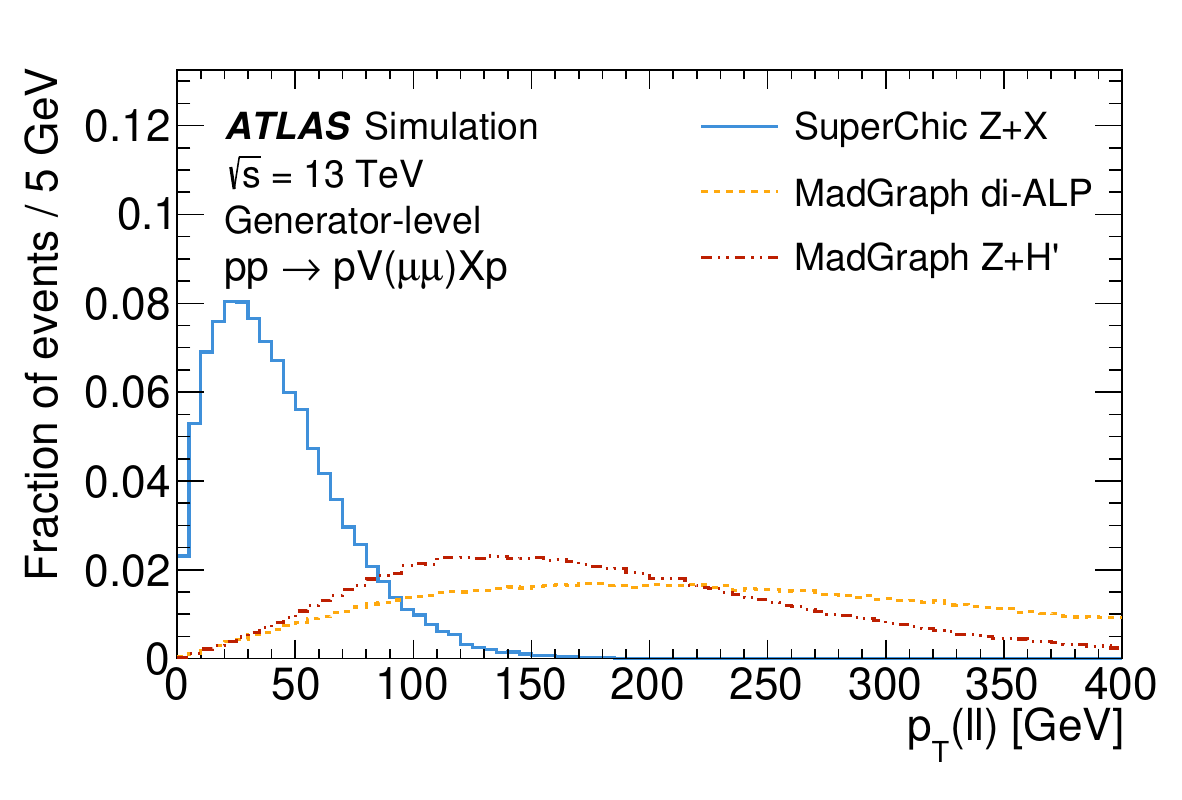}
\caption{}
\label{fig:sig_truth_comp_pTll}
\end{subfigure}
\begin{subfigure}{.45\textwidth}
\includegraphics[width=0.95\linewidth]{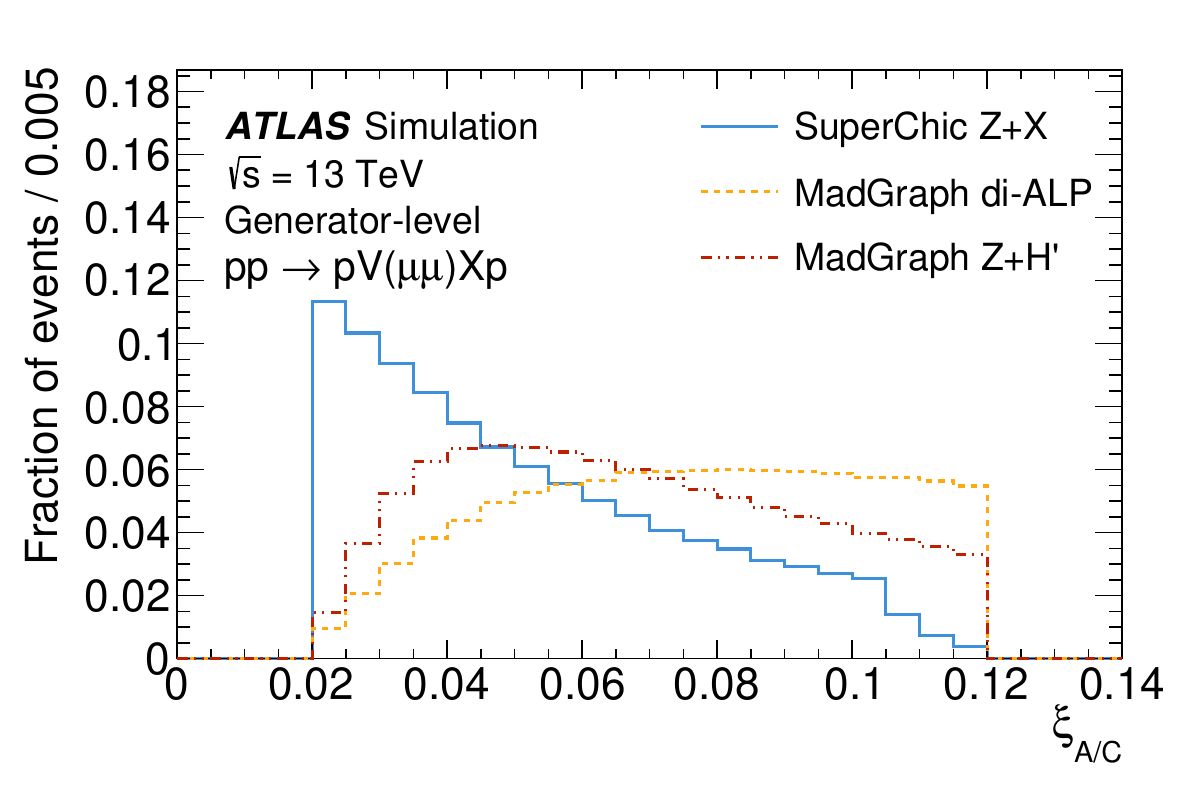}
\caption{}
\label{fig:sig_truth_comp_xi}
\end{subfigure}
\caption{Comparison of generator-level distributions of (a) dilepton $p_\textrm{T}$ and (b) proton $\xi$ for both signal protons, for each signal model with a hypothesised signal mass of $m_X = 500$~GeV. The plotted events are selected within the AFP acceptance of $0.02 < \xi < 0.12$.}
\label{fig:sig_truth_comp}
\end{figure}

\begin{figure}[ht]
\centering
\begin{subfigure}{.45\textwidth}
\includegraphics[width=0.95\linewidth]{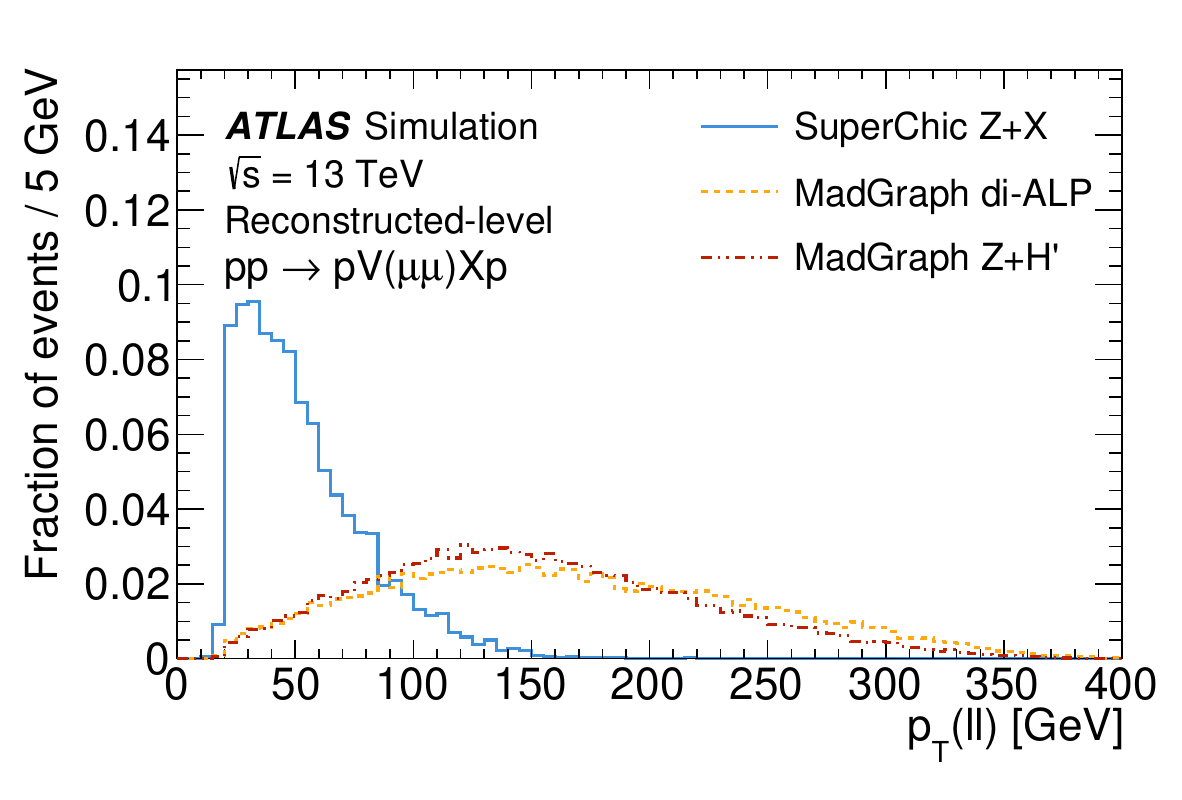}
\caption{}
\label{fig:signal model comp recon lepton pair pT}
\end{subfigure}
\begin{subfigure}{.45\textwidth}
\includegraphics[width=0.95\linewidth]{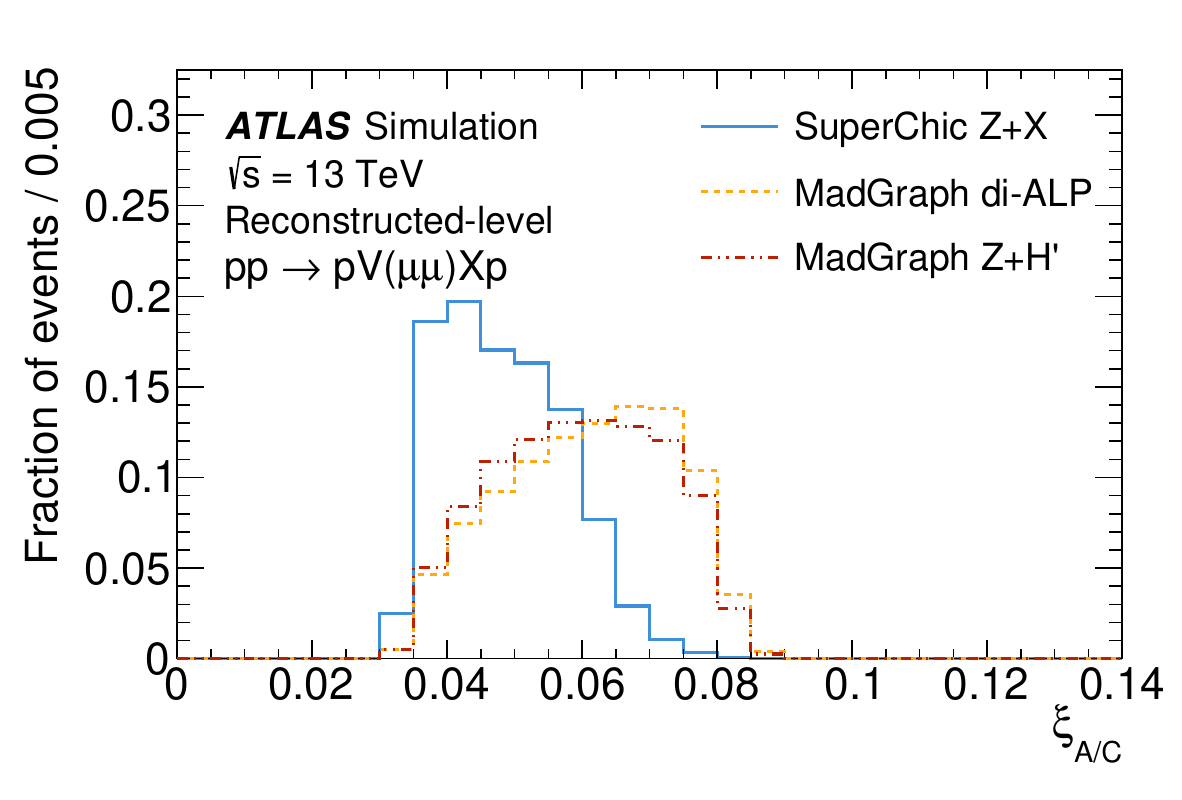}
\caption{}
\label{fig:signal model comp recon proton xi}
\end{subfigure}
\caption{Comparison of reconstruction-level distributions of (a) dilepton $p_\textrm{T}$ and (b) proton $\xi$ for both signal protons, for each signal model with a hypothesised signal mass of $m_X = 500$~GeV. The plotted events are selected within the AFP acceptance of $0.02 < \xi < 0.12$.}
\label{fig:signal model comp recon}
\end{figure}

As illustrated in Figure~\ref{fig:sig_truth_comp} for a representative signal mass of $m_X = 500$~GeV, the three models populate distinct kinematic regions at generator-level: the $Z+X$ model produces a less boosted final state, whereas both the $Z+H'$ and di-ALP models yield a more boosted final state. Comparing the $Z+H'$ and di-ALP models, the latter exhibits higher values of the dilepton transverse momentum and proton fractional energy loss.
Reconstruction-level distributions of the same kinematic variables are shown in Figure~\ref{fig:signal model comp recon}.

Owing to the wide kinematic spectrum of the $VX$ system and an associated effect on detector acceptance, particularly for the proton fractional energy loss $\xi$, a fiducial volume is defined. The volume definition is informed by the event selections applied to reconstructed events in this analysis. These are described in Section~\ref{sec:evnt_sel}, and represent the acceptance of the central ATLAS detector and AFP spectrometer. Fiducial selections are applied to the generator-level kinematics of the leptons and protons in simulated signal events, and are summarised in Table~\ref{tab:fiducial selection}.

\begin{table}[ht]
\caption{Summary of fiducial selection criteria for signal events.}
\centering
\begin{tabular}{lc}
\toprule
Feature & Criterion \\
\midrule\midrule
\multirow{2}{*}{Electrons} & $p_\textrm{T} > 18$~GeV \\
& $\left|\eta\right| < 2.47$ \\
\midrule
\multirow{2}{*}{Muons} & $p_\textrm{T} > 15$~GeV \\
& $\left|\eta\right| < 2.4$ \\
\midrule
\multirow{3}{*}{Dilepton system} & Same flavour, opposite charge \\
& $m_{\ell\ell} > 50$~GeV \\
& $p_\textrm{T}^{\ell\ell} > 20$~GeV \\
\midrule
Protons & $0.035 < \xi < 0.08$ \\
\bottomrule
\end{tabular}
\label{tab:fiducial selection}
\end{table}

The fiducial volume is used to define the reference normalisation of the signal. Signal events falling outside the fiducial region are treated as an additional source of background in the analysis.

\subsection{Simulated samples for background model validation}
The background in this analysis is almost exclusively combinatorial background, which results from the coincidence of a non-signal central dilepton signature and two uncorrelated protons that are deflected in different collisions in the same LHC bunch crossing (pile-up) and are detected in the AFP spectrometer. The two protons typically come from different pile-up interactions.
This background is fully modelled using a data-driven method called \enquote{event mixing}, described in Section~\ref{sec:background modelling}. However, in order to validate this method, MC event samples were generated for each of the contributing central processes and received a full $\GEANT$-based simulation of the ATLAS detector response.

The main process considered for the validation is inclusive Drell--Yan $Z$-boson production ($Z$+jets), which was simulated using \Sherpa[2.2.11]~\cite{Bothmann:2019yzt} for both parton generation and showering, with the \NNPDF[3.0NNLO] PDF set~\cite{Ball:2014uwa}. Inclusive top-quark production ($t\bar{t}$ and $Wt$) was simulated using \POWHEG[4]~\cite{Frixione:2007nw} with the \NNPDF[3.0NLO] PDF set, interfaced with \Pythia[8.244]~\cite{Sjostrand:2014zea} for parton showering with the A14 tune and the \NNPDF[2.3LO] PDF set \cite{Ball:2012cx}. Inclusive diboson production processes ($WW$, $WZ$, $ZZ$) were generated using \Sherpa[2.2.14] with the \NNPDF[3.0NNLO] PDF set. Exclusive samples of photon-induced lepton production and $WW$ boson production were generated using \MADGRAPH[2.9.5] with the CT14qed\_proton PDF set, with parton showering provided by \Pythia[8.245] with the A14 tune and \NNPDF[2.3QED] PDF set.

For all these central backgrounds with two final-state leptons, pile-up protons were overlaid from a database created from the 2017 dataset, to simulate the proton component of the combinatorial background.

\section{Event selection and object reconstruction}
\label{sec:evnt_sel}
The reconstruction-level event selection consists of a preselection and a final signal selection. The preselection imposes loose requirements to select events that have a signal-like topology. This provides a dataset which was used in various optimisation studies to help define the analysis strategy. The signal selection imposes additional requirements on top of the preselection, for optimal signal efficiency and background reduction, and events in the corresponding signal region are used in the final fit. %

Preselected events must contain at least one interaction vertex with two or more associated ID tracks that satisfy $p_\textrm{T} > 500$ MeV, $|\eta| < 2.5$, and additional loose quality criteria described in Refs.~\cite{ATL-PHYS-PUB-2015-051, PERF-2015-08}, corresponding to at least one pair of same-flavour leptons with opposite electric charge ($e^+e^-$ or $\mu^+\mu^-$). Electrons (muons) must satisfy the kinematic requirements $p_\textrm{T} > 18~(15)$~GeV and $|\eta| < 2.47~(2.4)$, in addition to loose identification criteria~\cite{EGAM-2018-01, MUON-2018-03} and $\left|z_0 \sin \theta \right| < 0.5$~mm.\footnote{$z_0$ is the longitudinal impact parameter relative to the primary vertex, which is defined as the vertex with the largest $\sum p_\textrm{T}^2$ of associated tracks.} Electrons that share an ID track with a muon are removed, unless the muon is of low quality (i.e.\ reconstructed without the muon spectrometer), in which case the muon is removed. If multiple electron candidates share a track, only the highest-$p_\textrm{T}$ candidate is kept. To reduce backgrounds from non-prompt or misidentified leptons, the remaining electrons (muons) must satisfy a transverse impact parameter significance $\left|d_0/\sigma_{d_0} \right| < 5 (3)$, in addition to the loose isolation criteria described in Refs.~\cite{EGAM-2018-01, MUON-2018-03}, which require there to be no additional particles above a given $p_\textrm{T}$ threshold within a variable-radius cone around each lepton track. In simulated event samples, scale factors are applied to leptons to match the reconstruction and trigger efficiencies measured in data~\cite{EGAM-2018-01, MUON-2018-03}. Protons must pass the \enquote{loose} selection, defined to match the AFP spectrometer acceptance of $0.02 \leq \xi \leq 0.12$.

Further lepton selection requirements are applied to the preselected events to define the signal region used in the final fits. Electrons (muons) must satisfy the \enquote{LooseAndBLayer}~\cite{EGAM-2018-01} (\enquote{Medium}~\cite{MUON-2018-03}) identification requirements, and the dilepton invariant mass must satisfy $m_{\ell \ell} > 50$~GeV. In addition, based on the higher dilepton transverse momentum observed in simulated signal events than in data events, a requirement of $p_\textrm{T}^{\ell \ell} > 20$~GeV is imposed on the dilepton system to enhance the signal-to-background ratio.

Finally, signal events are rejected if they have any tracks in the ID within a window of $|z_0^\textrm{track} - z_0^{\ell\ell}| < 0.5$~mm, where $z_0^\textrm{track}$ is the track $z_0$ position and the signal leptons $\ell_{1,2}$ form a dilepton vertex with $z_0^{\ell\ell} = (z_0^{\ell_1} + z_0^{\ell_2})/2$.
Considered tracks must satisfy the standard selection of $p_\textrm{T} > 500$~MeV, $|\eta| < 2.5$, and the loose quality criteria, in addition to a requirement of $|d_0| < 0.5$~mm, which is applied to reduce the rate of misidentified tracks. The signal lepton tracks are excluded from the track counting, which is achieved using direct ID matching for muon tracks, and a $\Delta R(\text{track}, \ell) < 0.01$ requirement for electron tracks. The event rejection described above is called the track veto ($N_\textrm{tracks}^\textrm{0.5~mm} = 0$), and is highly efficient in removing background events with inner-detector activity in addition to the dilepton vertex.

\begin{figure}[ht]
\centering
\includegraphics[width=0.7\linewidth]{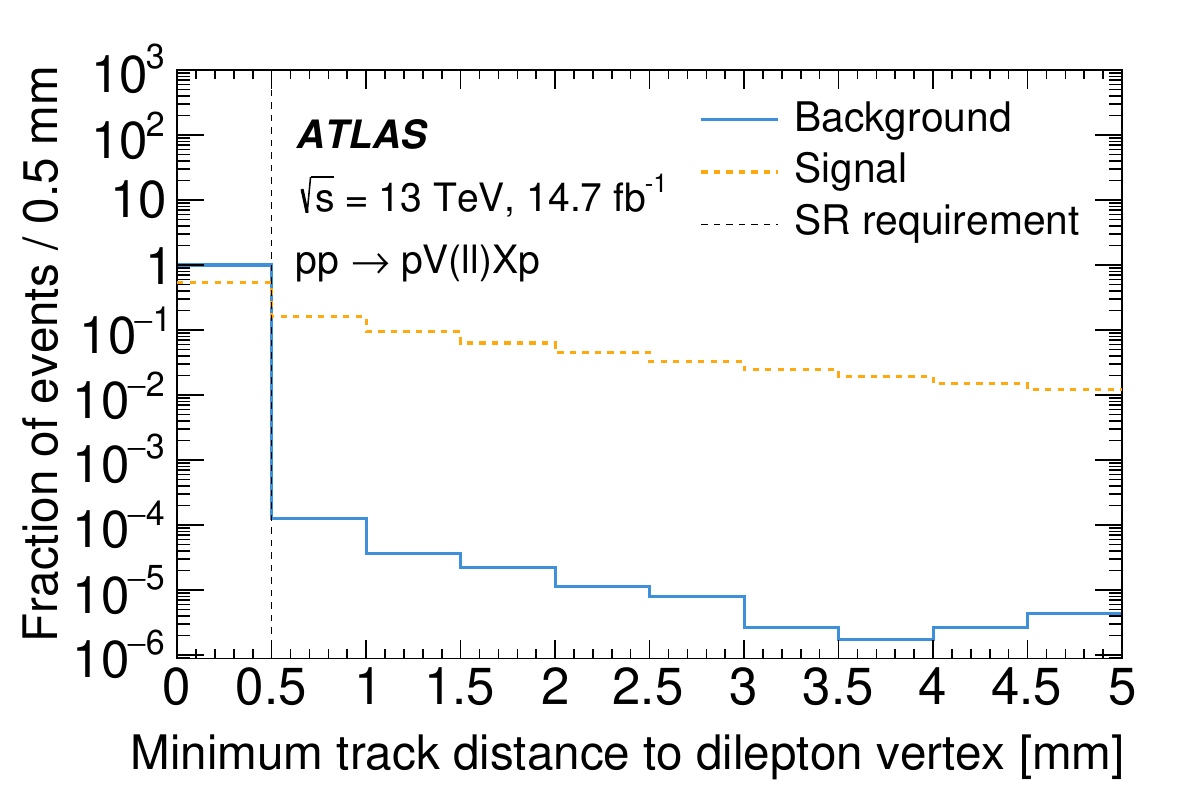}
\caption{Normalised distributions of the minimum $z$-axis distance between the dilepton vertex and additional tracks in background and signal events. For background events this is evaluated directly using the data-driven background model described in Section~\ref{sec:background modelling}, while for signal events it is taken from a pile-up-based method using the primary vertex position in the previous event. The distributions from all generated signal samples are summed since they are independent of the underlying model or signal mass. The signal selection requires there to be no additional tracks within 0.5~mm of the dilepton vertex.}
\label{fig:track veto effect signal vs background}
\end{figure}

Figure~\ref{fig:track veto effect signal vs background} shows the distribution of the $z$-axis distance between the dilepton vertex and the closest additional track in each event, for signal and background processes.
In background events, additional tracks are expected near the dilepton vertex due to additional event activity, such as jet production. This distribution is estimated relative to the dilepton vertex in preselected data events, almost all of which are expected to come from background processes. The signal process does not produce visible ID event activity other than the dilepton system, so additional tracks are not expected near the dilepton vertex. However, the track veto removes some signal events that have random coincidences with tracks from independent pile-up interactions close to the dilepton vertex.
This is also estimated from data, in this case by identifying which additional track in a given event is closest to the primary vertex position in the previous event, and building the distribution of distances between the two. This is referred to as the \enquote{pile-up-based method}. By measuring a track position that is independent of the dilepton vertex of the current event, it allows the probability of a pile-up track to randomly intercept the dilepton vertex to be estimated. Sampling the tested position by using the primary vertex of the previous event ensures that the distribution of tested track positions matches the expected distribution of dilepton vertex positions. A window size of 0.5~mm around the dilepton vertex was chosen for the track veto after optimisation with respect to signal sensitivity.

The preselection requires exactly one \enquote{loose} reconstructed proton (satisfying $0.02 \leq \xi \leq 0.12$) per AFP side or exactly one \enquote{tight} proton ($0.035 \leq \xi \leq 0.08$) per AFP side, for $m_X$ to be calculated. Events with more than one proton detected on either side, due to additional protons from pile-up interactions, cannot be used because it is not possible to tell which proton originated in the primary interaction and is thus associated with activity in the central region. For the signal region, only tight protons are accepted, corresponding to the $\xi$ region where the detector efficiency is well understood, and exactly one tight proton is required on each side. The \enquote{or} condition in the preselection is needed because requiring only exactly one loose proton per side would not fully contain the signal region. This can be understood by considering an event with a single proton with $\xi = 0.05$ on one AFP side and two protons with $\xi = 0.06$ and $0.1$ on the other. The two protons satisfying loose selection on one side would cause this event to fail the selection requiring exactly one loose proton per side. However, only one of these protons satisfies the tight selection, so this event would be accepted in the signal region. Including the \enquote{or} condition prevents such signal events from being excluded from the preselection. Additionally, the preselection accepts both single-station and double-station reconstructed protons, while the signal region requires double-station reconstruction.

In the final fits discussed in Section~\ref{sec:result}, an additional restriction is imposed on the missing mass $m_X$. A window is defined for the signal region, relative to whichever hypothesised signal mass is being considered. These are defined in Section~\ref{sec:result} for each fit.

\section{Background modelling}
\label{sec:background modelling}
Several sources of background are considered in the analysis:
\begin{itemize}
\item Inclusive SM processes ($Z$+jets, $t\bar{t}$, $Wt$, dibosons) with two protons from pile-up events (combinatorial background).
\item Exclusive SM processes (e.g.\ $\gamma\gamma \rightarrow \ell\ell$, $\gamma \gamma \rightarrow W W$).
\item Single- or double-dissociative $VX$ events.
\end{itemize}
The dominant source is the combinatorial background, which arises when central dilepton systems produced by non-signal SM processes are wrongly combined in reconstruction with unassociated protons originating in independent pile-up interactions. Each background proton typically comes from a different pile-up interaction, while the central component originates from one of the inclusive SM processes listed above. Exclusive SM processes with intact protons in the final state act as an additional background in this analysis, but it was determined from simulation that the participating protons typically fall outside the $\xi$ acceptance of AFP spectrometer because of the relatively low mass of the central system in these processes. Therefore, the exclusive central system is instead usually reconstructed along with pile-up protons as in the combinatorial background, and so can be modelled in the same way.

The combinatorial background is modelled using a data-driven technique known as event mixing. In this procedure, central-detector information (leptons and tracks) from a given real preselected data event is combined with AFP information (protons) from a different preselected event, shifted by an offset $i$ along the sample of preselected events. The offset $i$ is called the event-shift. The resulting mixed sample accurately mimics the characteristics of the combinatorial background. The shift index $i$ can take any value from $1$ to the total number of events, $N$, yielding up to $N$ uncorrelated mixed samples to model the background (with $i=0$ giving the unaltered data). For example, in a dataset with 4 events, there are 4 of each component: Central 1, 2, 3 and 4 and AFP 1, 2, 3 and 4. In normal data reconstruction, the components Central 1 and AFP 1 are combined to reconstruct $X$. However, when the event-mixing technique is applied, an event-shift of $i = 2$, for example, means that Central 1 is instead combined with AFP 3 (event number in the dataset shifted by the value of $i$), then Central 2 is combined with AFP 4, Central 3 with AFP 1 and Central 4 with AFP 2. In this way, 4 reconstructed $X$ systems are obtained with the two components mixed up, which gives a very accurate model of the dominant combinatorial background process. Within this construction, multiple independent background models can be created using different values of event-shift $i$, since the event components used to reconstruct the $X$ system are different in each sample. In this analysis, $100$ such samples were generated and averaged bin-by-bin to produce a high-statistics, data-driven background model, using event-shifts in the range $2\leqslant i \leqslant 101$ ($i=1$ was reserved for studies during the analysis). This model is then used in the fits described in Section~\ref{sec:result} to obtain the final results. The event-mixing procedure has been used previously by several analyses using forward proton information~\cite{STDM-2018-16, EXOT-2019-28, CMS-EXO-19-009}.

\begin{figure}[h!]
\centering
\begin{subfigure}{.45\textwidth}
\includegraphics[width=0.95\linewidth]{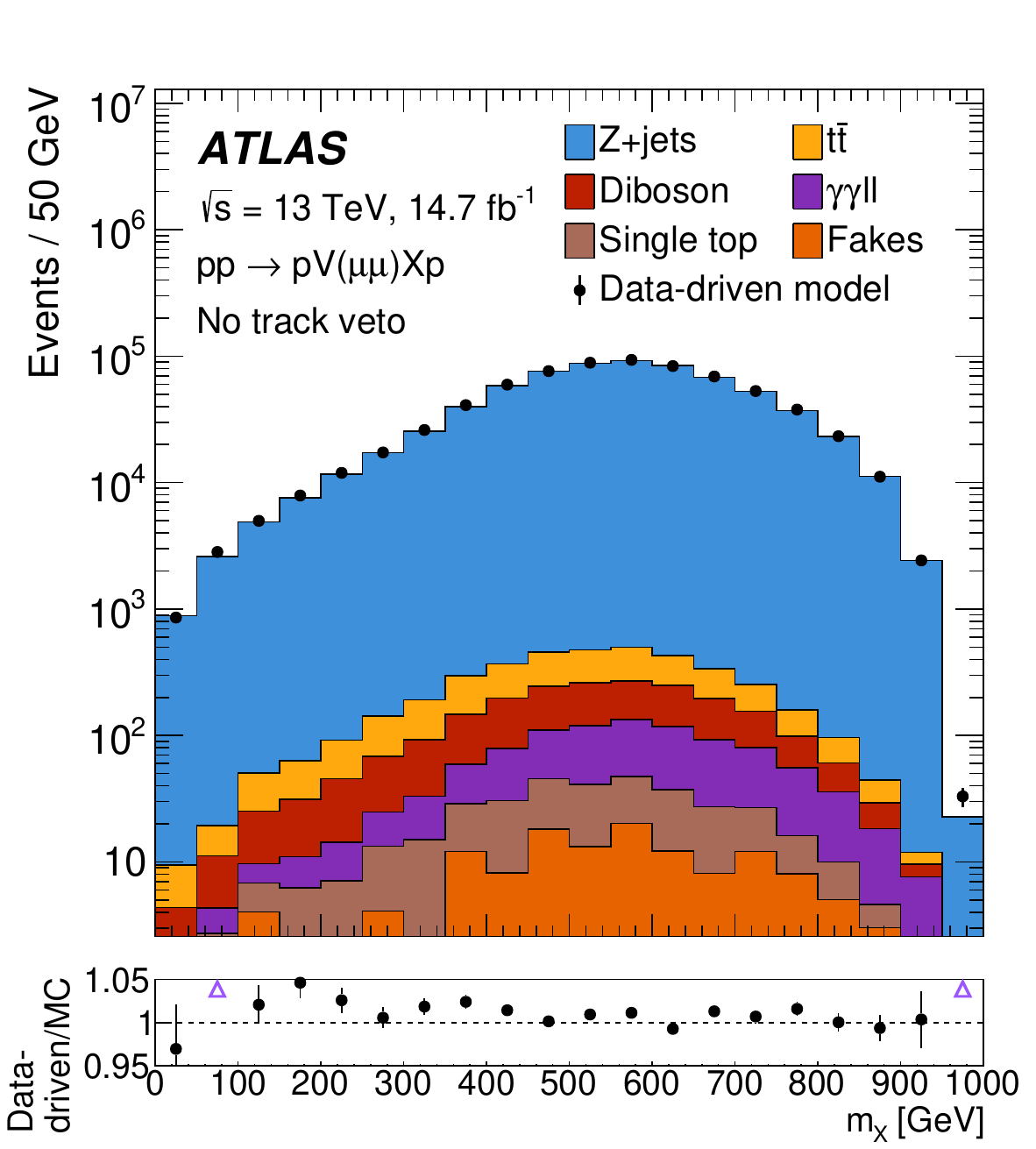}
\caption{}
\label{fig:multi bg comb mx before veto muons}
\end{subfigure}
\begin{subfigure}{.45\textwidth}
\includegraphics[width=0.95\linewidth]{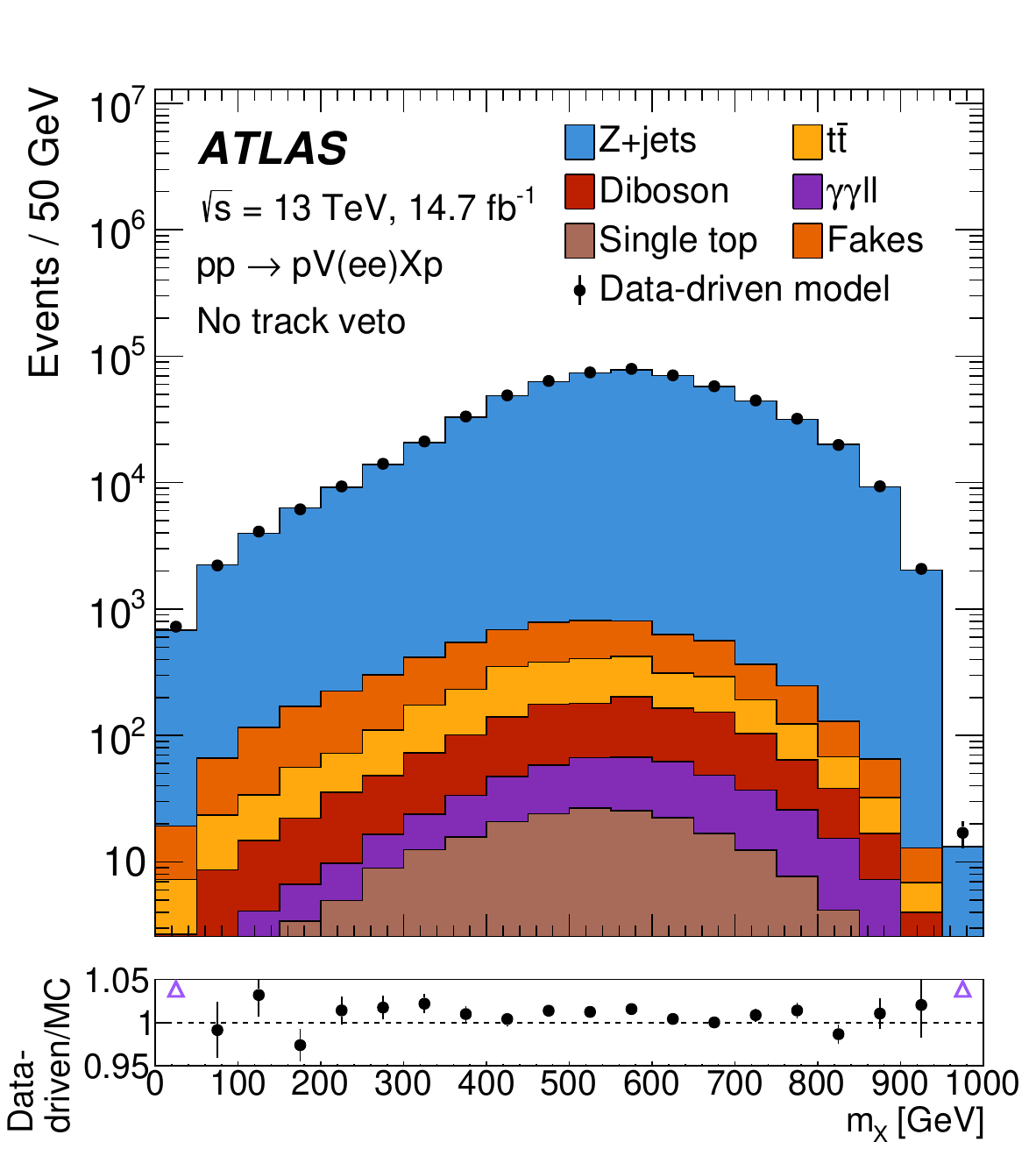}
\caption{}
\label{fig:multi bg comb mx before veto electrons}
\end{subfigure}\\
\begin{subfigure}{.45\textwidth}
\includegraphics[width=0.95\linewidth]{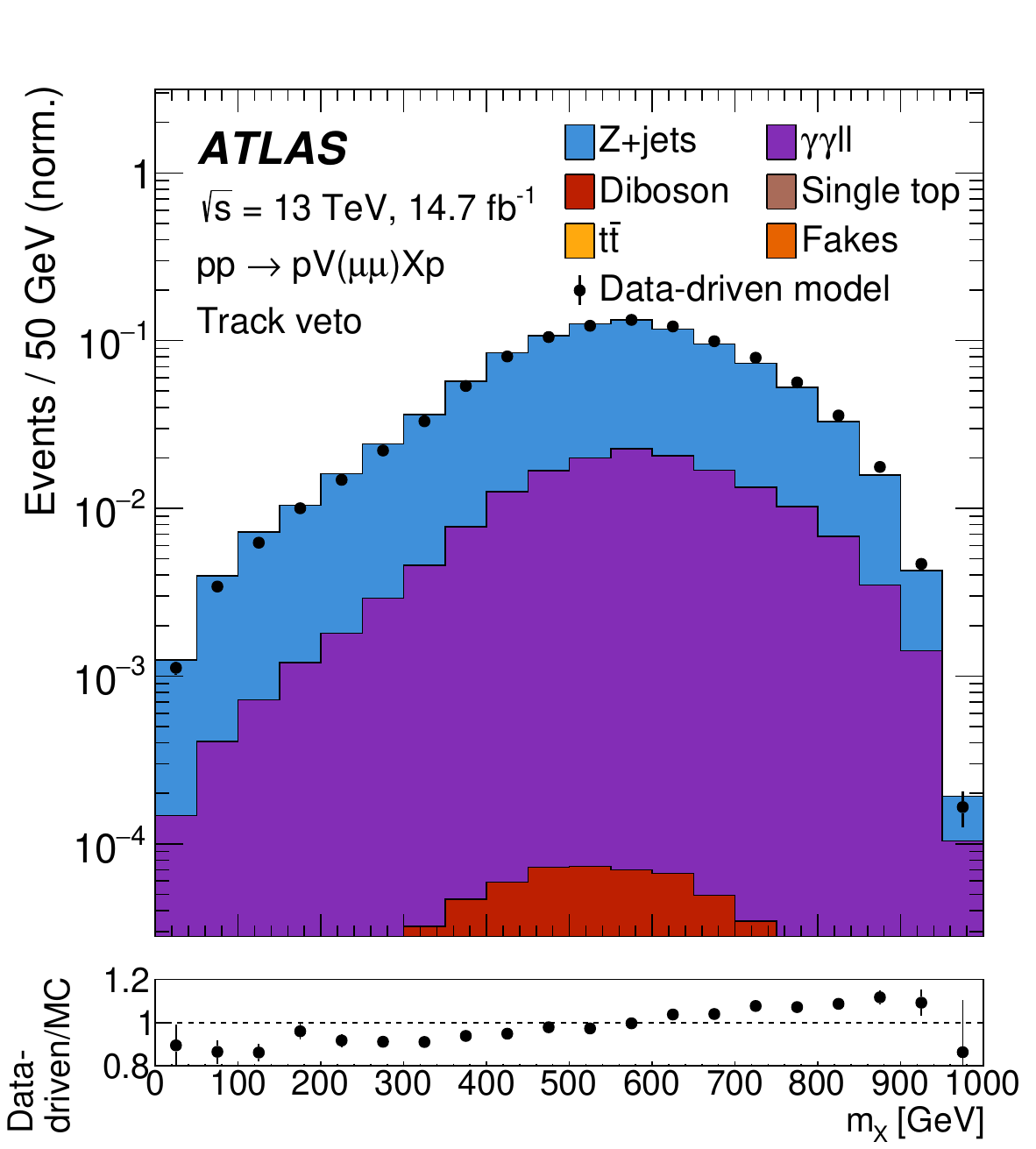}
\caption{}
\label{fig:multi bg comb mx after veto muons}
\end{subfigure}
\begin{subfigure}{.45\textwidth}
\includegraphics[width=0.95\linewidth]{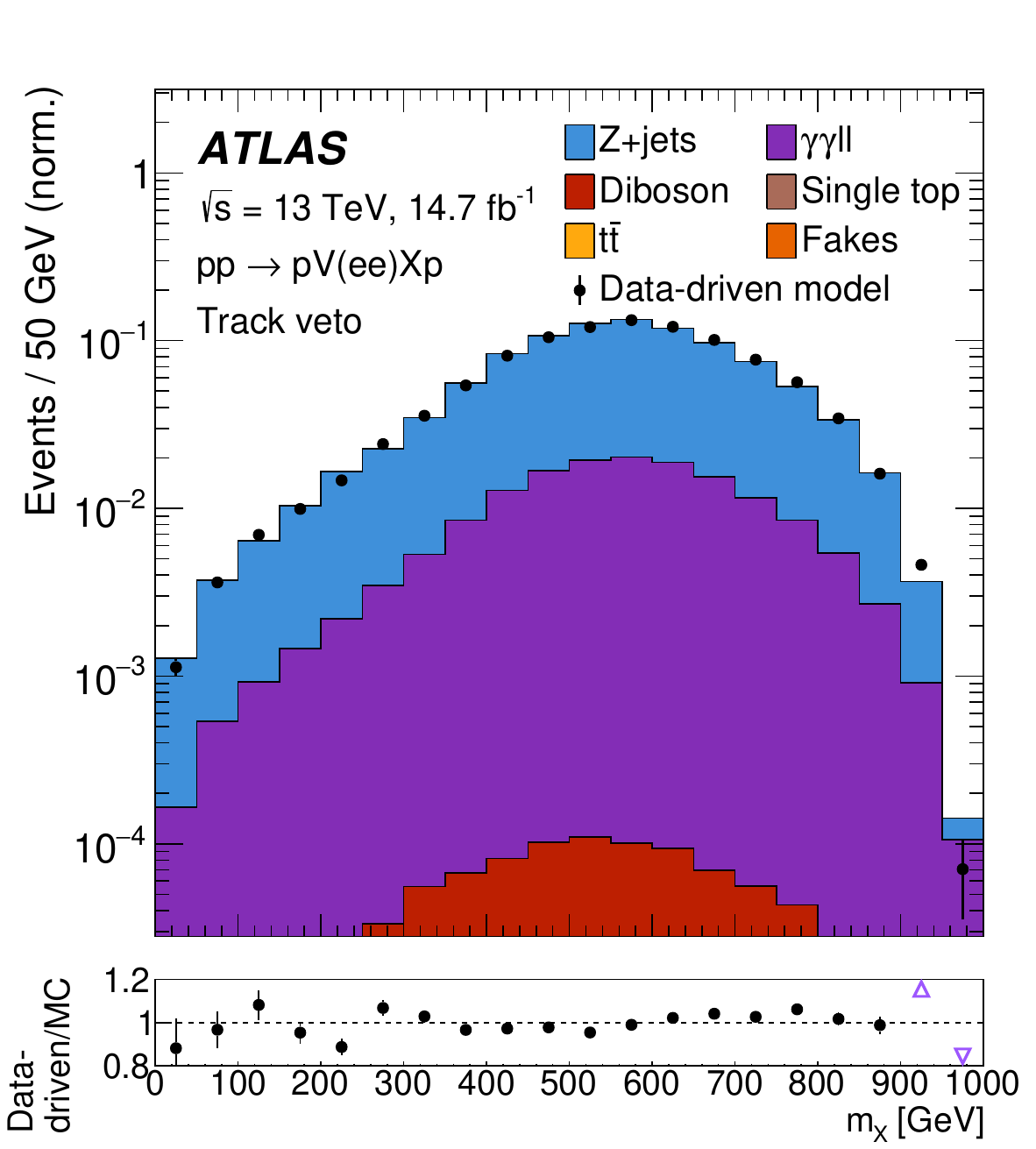}
\caption{}
\label{fig:multi bg comb mx after veto electrons}
\end{subfigure}
\caption{Missing-mass distributions in the overall simulated background model produced with all considered background contributions, after the final signal selection is applied except for the track veto, in the (a) muon and (b) electron channels and after the track veto is applied in the (c) muon and (d) electron channels. The data-driven background model is overlaid, showing good agreement with the simulated model. Distributions shown after applying the track veto are normalised to unity, to remove dependence on the known mismodelling of the underlying event in simulated event samples, which was found to yield inaccurate estimates of the track veto background efficiency. Exclusive $\gamma\gamma\rightarrow WW$ production was also investigated, but made a negligible contribution in all cases. No fiducial selection is applied to simulated background samples.}
\label{fig:multi bg comb mX}
\end{figure}

In order to validate this method, the data-driven combinatorial background model was compared against simulated background MC samples, as described in Section~\ref{sec:data_mc}. These samples include the relevant central-detector processes which give a dilepton signature similar to the signal. Pile-up protons from 2017 data were overlaid onto the inclusive MC events, thereby reproducing the full final state. Figure~\ref{fig:multi bg comb mX} shows a comparison of the missing-mass distributions obtained from the two background estimation methods (data-driven event-mixed and MC simulation) using the full signal selection, both with and without the track veto applied, for each lepton channel.
The contribution from most backgrounds, including misidentified (\enquote{fake}) leptons, is negligible once the track-veto requirement is applied. It heavily suppresses the $Z+$jets background, which nevertheless remains the dominant background source. In contrast, the exclusive $\gamma \gamma \rightarrow \ell \ell$ background becomes more prominent, as its exclusive nature makes it almost completely unaffected by the track veto. Small normalisation and shape differences of the order of 10\% are observed following the application of the track veto; however, this is considered to be due to insufficient precision in the modelling of the underlying event, as explained in the next paragraph, and does not suggest an issue with the data-driven background.

For this comparison, distributions are normalised after applying the track veto. This is necessary because the description of the underlying event in Drell--Yan processes has been observed to be insufficiently precise for the estimation in the low $N_\textrm{track}$ region, as shown in Figure~3 of Ref.~\cite{STDM-2017-21} and in Refs.~\cite{STDM-2014-13, STDM-2011-42}. Several background MC generators were tested in this regard, and found to overestimate the fraction of events passing the track veto by factors that can be large, consistent with those found in a previous ATLAS analysis~\cite{STDM-2017-21}. This emphasises the need to use a data-driven background in this analysis, as there is no dependence on modelling and thus no effect from associated issues such as the underlying event.

To avoid bias from potential signal contamination in the data-driven background model, the normalisation is obtained from a background-only fit in a control region where negligible signal is expected, as explained in more detail in Section~\ref{sec:result}.

Potential bias from repeated central components for several events in the data-driven background model is not considered to affect the model significantly, because the event-mixing procedure uses the preselected dataset of about $3\times 10^6$ events, and only about $2\times 10^5$ events remain after applying the signal selection (except for the track veto). Therefore, even after averaging over 100 different event-mixed samples, no more than ${\sim}8$ repetitions would be expected in the final model, and in each case the reconstructed missing mass would be greatly varied due to combination with a different set of protons.

An additional signal-induced background component arises from events where at least one of the signal protons is not detected by the AFP spectrometer, and in its place a single pile-up proton is detected. This occurs mainly due to the signal proton falling outside the $\xi$ acceptance of the AFP spectrometer, and less often due to dissociation of the signal proton following the central interaction. Due to the event selection procedure, such an event would be accepted as signal and reconstructed, but since the energy loss $\xi$ is generally very different from that of the signal proton, the reconstructed missing mass would be inaccurate. Typically, this occurs when the signal proton has an energy loss significantly higher than the upper limit of AFP acceptance, resulting in a highly energetic dilepton pair which is not balanced in Eq.~(\ref{eqn:missing mass calculation}) by the pile-up proton which is reconstructed instead. This can result in negative reconstructed missing mass, which is unphysical and thus easily removed from consideration. However, if the missing mass is positive then such events cannot be easily distinguished from genuine signal events. These \enquote{mismatched} signal events form a wide resonance resembling the combinatorial background, beneath the genuine signal peak. Some weak signal-mass dependence is seen: this effect is most significant for low- and high-mass signal models on the edge of the acceptance in this analysis, where in addition to the genuine signal occurring at the centre of the $m_X$ distribution there are large contributions away from the signal peak.

\begin{figure}[h!]
\centering
\begin{subfigure}{.45\textwidth}
\includegraphics[width=0.95\linewidth]{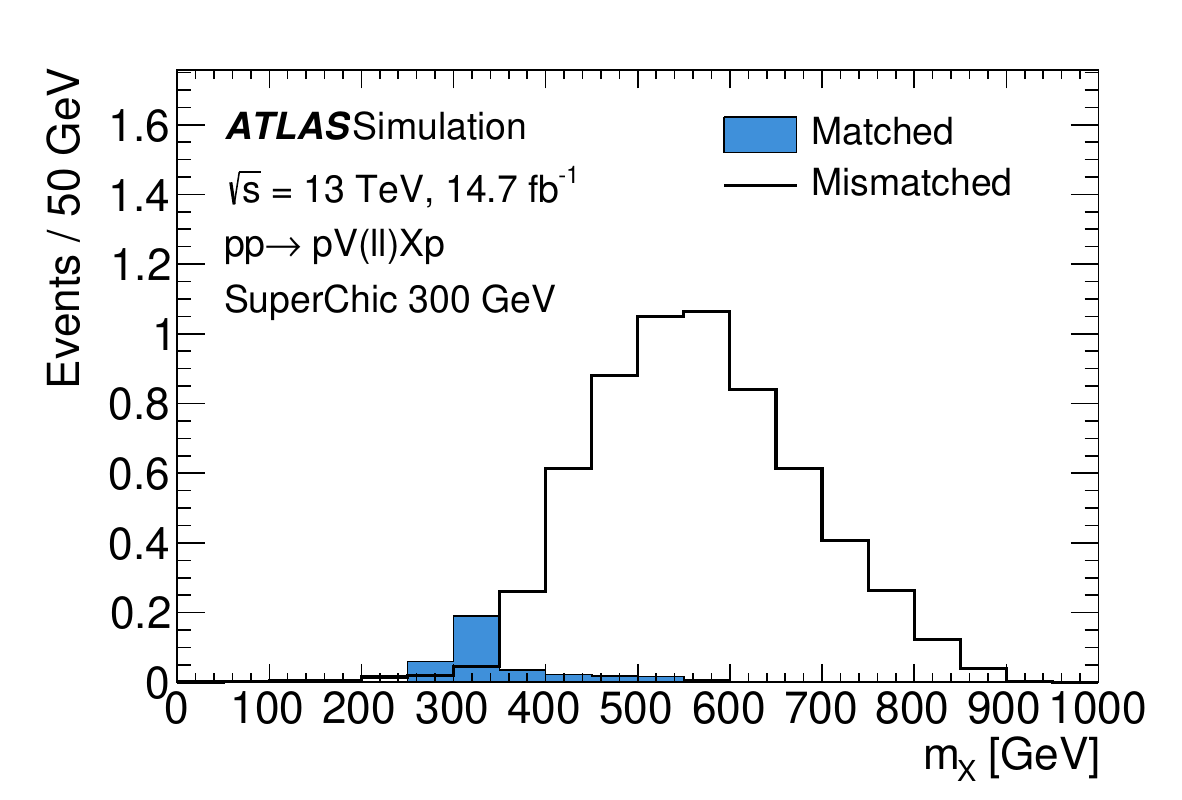}
\caption{}
\label{fig:match vs mismatch sc 300 no fid sel}
\end{subfigure}
\begin{subfigure}{.45\textwidth}
\includegraphics[width=0.95\linewidth]{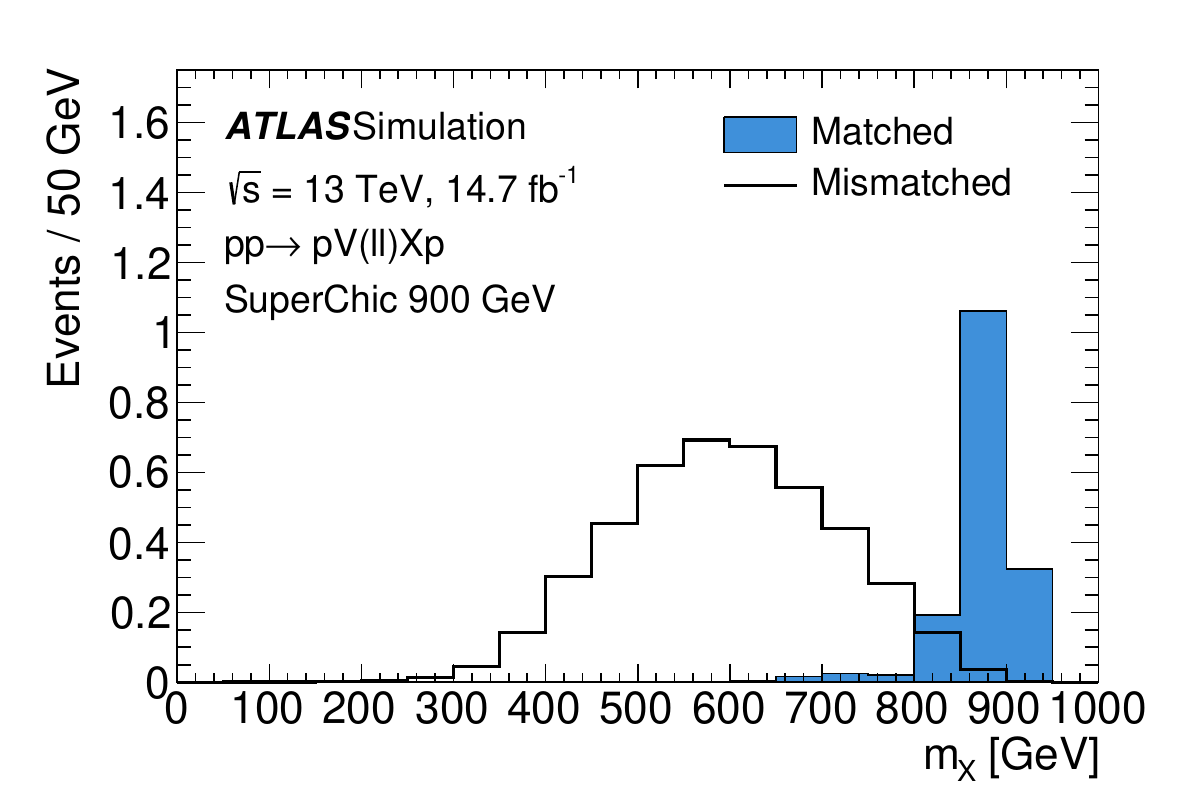}
\caption{}
\label{fig:match vs mismatch sc 900 no fid sel}
\end{subfigure}
\caption{Comparison of missing-mass distributions for simulated signal events with both reconstructed protons matched at generator-level to the signal process, and with at least one originating in pile-up (mismatched), for the \textsc{SuperChic} signal model at hypothesised signal masses of (a) 300~GeV and (b) 900~GeV without the generator-level fiducial selection applied.}
\label{fig:match vs mismatch}
\end{figure}

Figure~\ref{fig:match vs mismatch} shows a comparison between signal events with \enquote{matched} and \enquote{mismatched} protons for two \textsc{SuperChic} models with signal masses of $m_X = 300$~GeV and 900~GeV.
Distributions are shown before the generator-level fiducial selection is applied to the simulated signal, to demonstrate both components which would be present in data which contained signal events. The large contribution from the mismatched component is clearly visible, as these masses are at the tails of the $m_X$ distribution and therefore most heavily affected by signal protons falling outside AFP acceptance. As discussed in Section \ref{sec:data_mc}, the fiducial selection is used to separate simulated signal events into two categories: inside and outside of the fiducial volume. Events inside the fiducial volume are treated as signal, and events outside the fiducial volume are treated as an additional background component. The majority of mismatched signal events discussed above fall outside the fiducial volume, due to the generator-level $\xi$ selection requiring both signal protons to fall within the AFP spectrometer acceptance. As a result, the simulated signal model used in the final fit (inside the fiducial volume) is highly pure in events where the genuine signal protons are reconstructed by AFP.
The background contribution from signal events outside the fiducial selection is strongly suppressed within the signal region by the mass window applied in the final fits, since for the most affected signal models the mismatched peak occurs away from the main signal peak, as shown in Figure~\ref{fig:match vs mismatch}. The normalisation of this background is fixed by the signal model investigated, and due to the mass window requirement is consistently negligible in the final distributions.

\section{Systematic uncertainties}
\label{sec:sys_unct}
A likelihood fit is used in the analysis to fit the missing-mass distribution predicted by the combined signal and background model to the observed data and determine the significance of the observed signal. Systematic uncertainties are incorporated into the likelihood fit as nuisance parameters (NPs), which affect the final distributions of the observables. These NPs are typically constrained by Gaussian priors reflecting the size of the uncertainties. They are profiled during the fit to extract the best-fit values of the parameters of interest. The instrumental uncertainties considered include systematic effects related to the reconstruction and calibration of physics objects used in the analysis, such as lepton momentum calibration and smearing, lepton reconstruction efficiencies, and lepton trigger inefficiencies. Additionally, the normalisation of simulated event samples to the integrated luminosity carries an uncertainty of $1.2\%$~\cite{DAPR-2021-01}, and an uncertainty is estimated for the reweighting of the pile-up distribution in simulated event samples to match the distribution observed in data.

\begin{figure}[ht]
\centering
\begin{subfigure}{.45\textwidth}
\includegraphics[width=\linewidth]{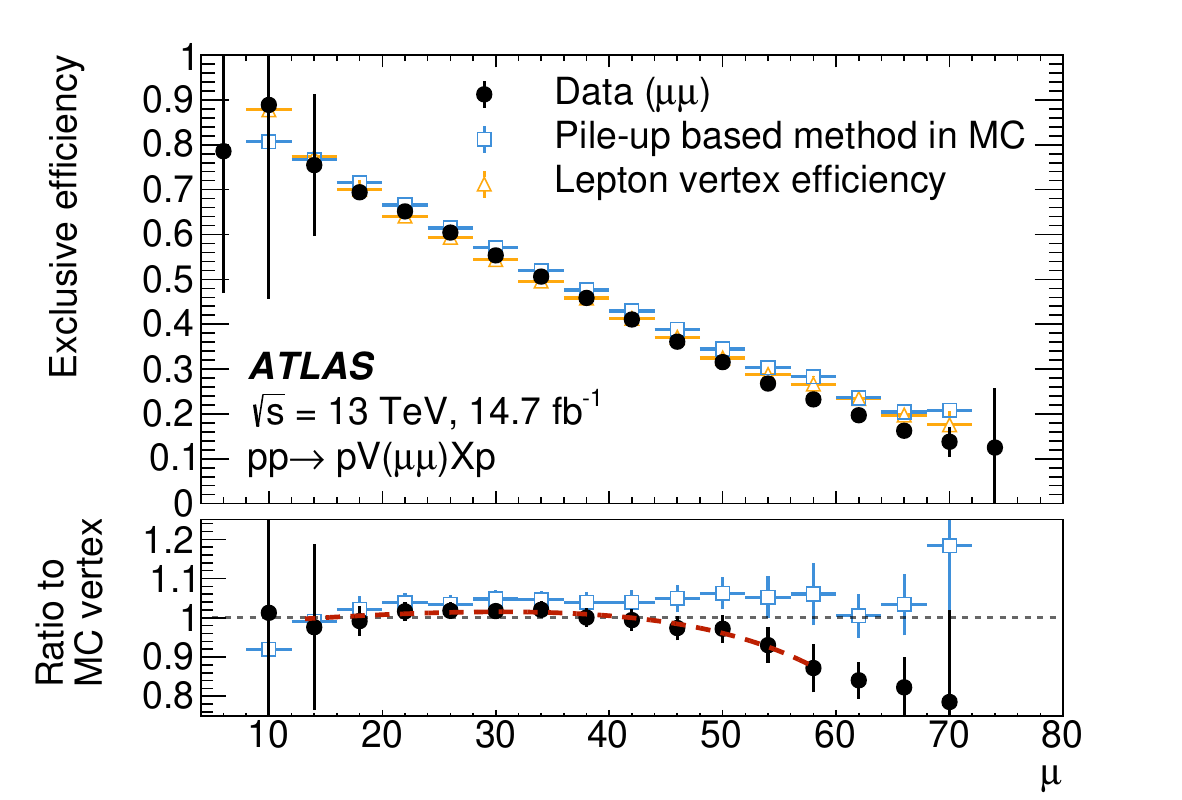}
\caption{}
\label{fig:track veto sig eff closure high pt muons}
\end{subfigure}
\begin{subfigure}{.45\textwidth}
\includegraphics[width=\linewidth]{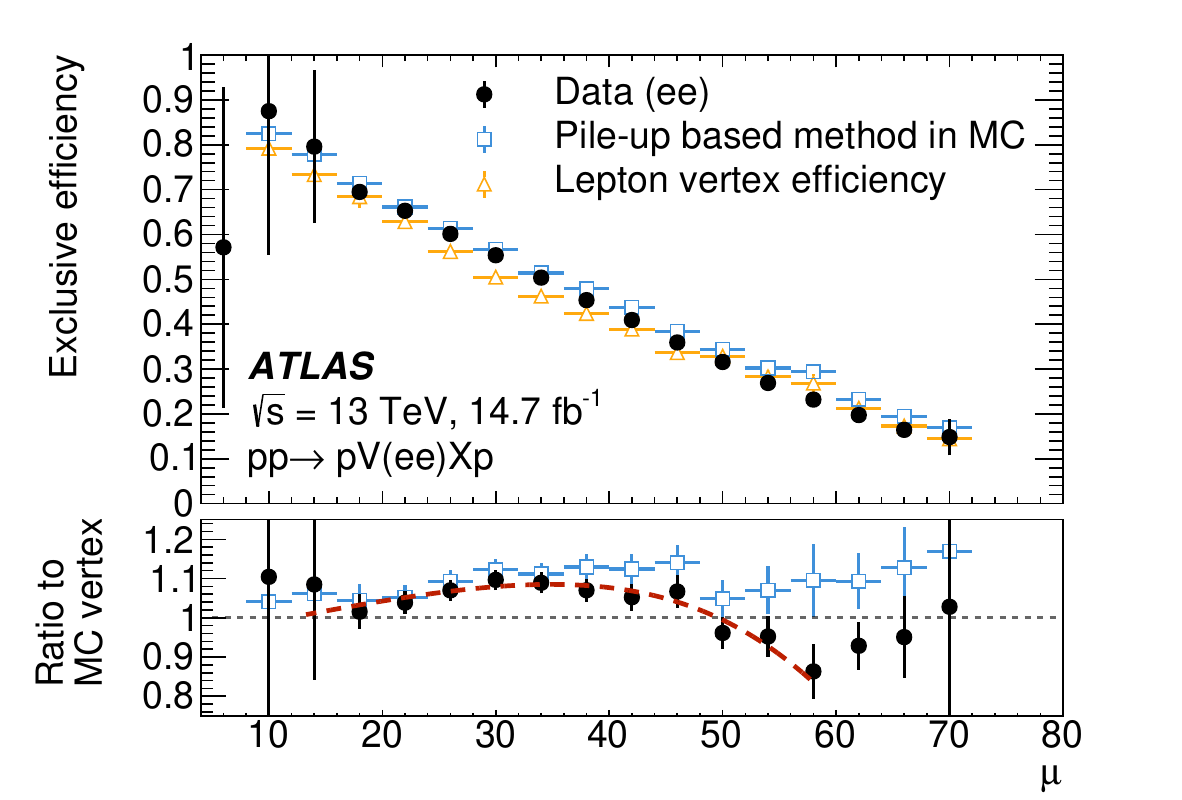}
\caption{}
\label{fig:track veto sig eff closure high pt electrons}
\end{subfigure}
\caption{Comparison of track-veto signal efficiencies estimated as a function of the mean number of interactions per bunch crossing, $\mu$, using a pile-up-based approach in simulated and real data and calculated directly from the dilepton vertex in simulated data in the (a) muon and (b) electron channels. The ratio of the pile-up-based estimate from data to the direct measurement in simulation is fitted to a quadratic polynomial, shown by the red dotted line, to remove dependence on statistical fluctuations. The fit range is limited to $\pm2\sigma$ around the mean value of $\mu$, where $\sigma$ denotes the standard deviation, to remove outliers. The ratio of efficiencies obtained by applying the two methods to simulation is consistent with the maximum ratio observed in data.}
\label{fig:track veto sig eff closure high pt}
\end{figure}

The uncertainty in the signal efficiency of the track veto is also evaluated. This efficiency is called the \enquote{exclusive efficiency}, since the inner-detector tracks in the track-veto window are required to be exclusively those of the two signal leptons. The uncertainty in this efficiency arises mainly from uncertainties in the modelling of pile-up interactions, and less significantly from the difference in luminous-region size between MC simulation, where it is fixed, and data, where it varies across runs. This leads to slight track-veto signal efficiency differences between data and simulation. To determine the uncertainty, the track veto efficiency was estimated using two different methods and compared as a function of the number of interactions per bunch crossing, with the largest observed difference being taken as the uncertainty. The first \enquote{direct} estimate is based on the fraction of simulated signal events which pass the veto. The second estimate is obtained using the \enquote{pile-up-based method} described in Section~\ref{sec:evnt_sel}, which measures the probability for pile-up tracks to fall within the veto window around the dilepton vertex by computing the minimum distance of tracks to an arbitrary $z$-axis position in each event. The second method was applied to both data and simulation. Each estimate was fitted using a quadratic function to mitigate sensitivity to outliers and statistical fluctuations, with the fit range limited to $\pm 2\sigma$ around the mean number of interactions per bunch crossing, $\mu$, observed in the data. The quadratic function was chosen as the lowest-order polynomial which consistently fitted the distributions with $\chi^2$ per degree of freedom close to 1. Figure~\ref{fig:track veto sig eff closure high pt} shows a comparison between the estimates in data and simulation and between the two different methods in simulation. The resulting uncertainties, estimated from the difference between the two methods, were found to be $12\%$ in the muon channel and $16\%$ in the electron channel. The larger uncertainty in the electron channel is attributed to Bremsstrahlung emission from signal electrons, in addition to larger effects from requirements on track-to-vertex matching, due to poorer $ee$ vertex resolution.
At the value of $\mu$ corresponding to the mean observed value in 2017 data ($\mu \sim 35$), the estimated track-veto signal efficiency obtained using the pile-up-based method on data is found to be $(50.6\pm0.3)\%$ in the muon channel and $(50.4\pm0.3)\%$ in the electron channel, where the uncertainties arise from limited statistics.

Systematic uncertainties associated with the proton reconstruction are also taken into account, as in previous analyses using AFP data~\cite{STDM-2018-16, EXOT-2019-28}. An uncertainty of $\pm300~\mu$m is assigned to the global alignment of the AFP spectrometer, along with a conservative uncertainty of $\pm0.05$~mm in the width of the Gaussian smearing used to calibrate the resolution of the reconstructed $\xi$ distribution of MC generator-level protons to match that observed in data. In addition, an uncertainty in the proton transport simulation used to determine the proton energy loss $\xi$ from the $x$ position measurement provided by the AFP spectrometer is accounted for by recalculating the proton properties using alternative transport simulations in which the beam angle is altered by $\pm50\,\mu$rad. The uncertainty in the measured proton reconstruction efficiency of $(92 \pm 2)\%$, which is applied to simulated event samples, was included as an additional systematic uncertainty. Furthermore, several variations affecting the proton reconstruction methodology (e.g.\ track finding and matching criteria) were considered, and it was verified that their impact on the data is negligible.

Modelling uncertainties affecting the data-driven background model are likewise considered. Statistical fluctuations in the background model are estimated using bootstrapping to be around 1\% due to the use of 100 statistically independent samples. Time dependence of the signal observable $m_X$ due to variations in the pile-up profile across the dataset is found to be negligible. The presence of single-diffractive backgrounds was considered by repeating the event-mixing procedure on only a single side of the AFP spectrometer, after first using the default double-sided mixing described in Section~\ref{sec:background modelling}, and revealed negligible shape differences. Finally, the effect of differences in the level of pile-up between events used for each component of the event-mixing procedure was found to have a negligible effect on the model.

Signal modelling uncertainties are associated with the choice of baseline set-up for the generation of signal events. As described in Section~\ref{sec:data_mc}, the \textsc{SuperChic} signal model uses an estimated scale factor to account for soft-survival effects. A soft-survival factor uncertainty of 20\% was applied to the signal normalisation in these samples. This was chosen as a conservative uncertainty based on several previous studies~\cite{STDM-2016-13,STDM-2018-16}, and does not reflect the intrinsic theoretical uncertainty of 1\% estimated in Ref.~\cite{Harland-Lang:2021ysd}. For \textsc{MadGraph} samples, the effect of soft-survival is not modelled by the generator, so the uncertainty for each sample is based only on a downward systematic variation that matches the scale factor estimated in \textsc{SuperChic} for the corresponding hypothesised signal mass, but capped at a minimum value of 20\% so as to not be lower than the uncertainty assigned to \textsc{SuperChic} models. This accounts for the overestimation of the signal yield obtained by neglecting soft-survival effects. Uncertainties related to parton shower settings were found to be negligible. Statistical uncertainties arising from the simulated signal samples are also considered.

\begin{table}[ht]
\caption{Summary of all considered uncertainties for the signal and background models, and their effects on the event yields of the corresponding samples. For the signal samples, each generator is shown separately, and the range of absolute percentage changes in event yield is given in a common signal mass range of 300--800~GeV for all models. Lepton uncertainties are combined for each channel, with a breakdown given only for proton uncertainties.}
\centering
\resizebox{0.87\linewidth}{!}{\begin{tabular}{lccc}
\toprule
\textbf{Uncertainty} & \multicolumn{3}{c}{\textbf{Effect on event yield}} \\
\midrule
\multicolumn{4}{c}{\textbf{Signal uncertainties}}  \\
\midrule
\textbf{Signal model} & \textsc{SuperChic} & \textsc{MadGraph} di-ALP & \textsc{MadGraph} $Z+H'$ \\
\midrule
Soft-survival factor & ~20\% & ~20\%--39\% & ~20\%--39\% \\
Track veto signal efficiency & ~16\% & ~16\% & ~16\% \\
Pile-up reweighting & 3.1\%--4.6\% & 3.6\%--4.4\% & 3.7\%--4.4\% \\
Luminosity & 1.2\% & 1.2\% & 1.2\% \\
\midrule
\textbf{Proton} (total) & ~~12\%--102\% & 22\%--60\% & 19\%--65\% \\
\midrule
AFP proton transport & 8.8\%--70\% & 16\%--43\% & 13\%--50\% \\
AFP global alignment & 5.7\%--63\% & 12\%--43\% & 11\%--54\% \\
AFP smearing & 4.8\%--38\% & 8.7\%--22\% & 7.6\%--25\% \\
AFP reconstruction efficiency & 3.2\%--3.7\% & 2.7\%--3.7\% & 3.4\%--3.7\% \\
AFP track matching & 1.2\%--2.5\% & 1.3\%--1.8\% & 1.3\%--1.6\% \\
AFP track finding & 0.6\%--1.2\% & <0.5\% & 0.5\% \\
\midrule
\textbf{Electron} (total) & 5.6\%--7.5\% & 5.5\%--8.2\% & 5.1\%--6.4\% \\
\midrule
\textbf{Muon} (total) & 5.5\%--7.4\% & 5.4\%--8.1\% & 5.0\%--8.0\% \\
\midrule
\textbf{Statistical uncertainty} & 1.1\%--3.4\% & 0.7\%--2.1\% & 0.5\%--1.8\% \\
\midrule
\textbf{Total} & ~~31\%--106\% & 30\%--75\% & 28\%--76\% \\
\midrule
\multicolumn{4}{c}{\textbf{Background uncertainties}} \\
\midrule
\textbf{Statistical uncertainty} & \multicolumn{3}{c}{1.0\%} \\
\bottomrule
\end{tabular}}
\label{tab:short syst summary}
\end{table}

Table~\ref{tab:short syst summary} presents a summary of the systematic uncertainties. The largest uncertainties are those affecting protons via the optics parameterisation and global alignment of the AFP spectrometer. Both of these cause the $\xi$ distribution and thus the $m_X$ distribution to shift upwards and downwards. This removes events on the edges of the $\xi$ acceptance, which has a significant effect on the event yield, especially for low and high mass points, which correspond to low and high $\xi$ values. This is particularly significant for the \textsc{SuperChic} model, in which the $Z$ and $X$ systems are produced with low $p_\textrm{T}$, meaning that the energy loss of the two interacting protons must be well balanced, imposing a relatively tight restriction on the possible $\xi$ values for a given hypothesised signal mass compared to the two \textsc{MadGraph} models. For low- or high-mass signal models, this restriction lies close to an acceptance edge, increasing the event yield uncertainty.

\section{Results}
\label{sec:result}

The results of the search are interpreted for each of the simplified models, and also as general limits on BSM physics cross-sections.

In each case, a search for the presence of a signal is performed using a profile-likelihood fit to the observed event yields in the missing-mass distribution, $m_X$. A single-bin fit is employed for each tested mass point, using a dedicated mass window centred on the hypothesised signal mass, as discussed in Section~\ref{sec:evnt_sel}. The resolution of the missing mass reconstructed using the method described in Section~\ref{sec:intro} was studied using simulated event samples, and found to be 50~GeV or better across the considered range of masses. Therefore, the chosen mass window extends 100~GeV to either side of the signal mass for most models (e.g.\ $400 \leq m_X \leq 600$~GeV for a 500~GeV signal model), although wider, asymmetric windows of $0 \leq m_X \leq 300$~GeV and $700 \leq m_X \leq 1000$~GeV are used for 100~GeV and 900~GeV models because there are fewer events in these regions in data. This provides an overlap of at least 100~GeV between adjacent tested models, which is more than the measured $m_X$ resolution of 50~GeV, ensuring that any resonances present within the tested range of signal masses will be visible.

The normalisation $\mu_{\text{bkg}}$ of the data-driven model of the combinatorial background is constrained using control regions. To compensate for potential bias in the background model from any signal contamination present in the dataset, the control regions are defined separately for each signal mass point. The mass windows defining the signal and control regions for the model-dependent results are given in Table~\ref{tab:SR and CR defs high-pT}.
\begin{table}[ht]
\centering
\caption{Signal and control region definitions used for the model-dependent results, with mass windows in the $m_X$ distribution.}
\begin{tabular}{c|c|c}
\toprule
Signal mass [GeV] & Signal region [GeV] & Control region(s) [GeV] \\
\midrule
100 & ~~~~0--300 & 300--400 \\
200 & 100--300 & 300--400 \\
300 & 200--400 & ~~~~0--100, 400--450 \\
400 & 300--500 & ~~600--1000 \\
500 & 400--600 & ~~700--1000 \\
600 & 500--700 & ~~400--450, 800--1000 \\
700 & 600--800 & ~~450--500, 900--1000 \\
800 & 700--900 & 400--600 \\
900 & ~~700--1000 & 400--600 \\
\bottomrule
\end{tabular}
\label{tab:SR and CR defs high-pT}
\end{table}

The control regions are defined as the range in $m_X$ outside the signal region, with a 100 GeV gap imposed on either side to ensure that no signal will enter the control region.
For some low signal mass points (100-300~GeV), the control region was extended into the gap region to recover high enough statistics to allow the fit to converge.
The CR window definitions were verified through several signal injection tests in which a large quantity of signal was injected into the data sample used to produce the data-driven background model. The observed limits obtained from normal data using this injected background were consistent within statistical uncertainty with the limits obtained using the normal background model, verifying that any signal contamination is effectively removed using this approach.

A simultaneous fit is performed to the signal and control regions with $\mu_{\text{bkg}}$ and $\mu_{\text{sig}}$ as free parameters, where $\mu_{\text{sig}}$ scales the signal normalisation relative to the pre-fit value and is allowed to be negative. In order to control the number of NPs and avoid overfitting, systematic uncertainty pruning is applied using a threshold of $0.5\%$ impact on the normalisation of the fitted distribution.

\begin{figure}[h!]
\centering
\begin{subfigure}{.45\textwidth}
\includegraphics[width=0.95\linewidth]{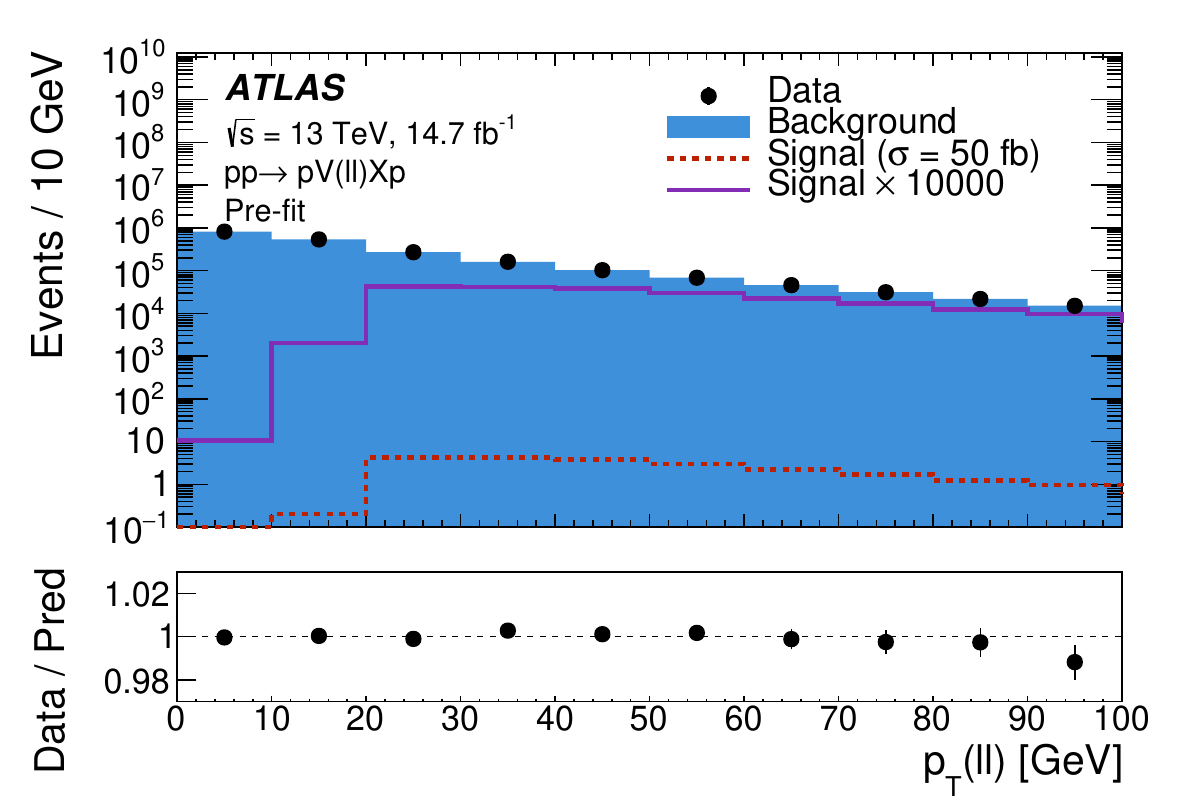}
\caption{}
\label{fig:unblinded data/bkg comp pre-sel lepton pair pt comb}
\end{subfigure}
\begin{subfigure}{.45\textwidth}
\includegraphics[width=0.95\linewidth]{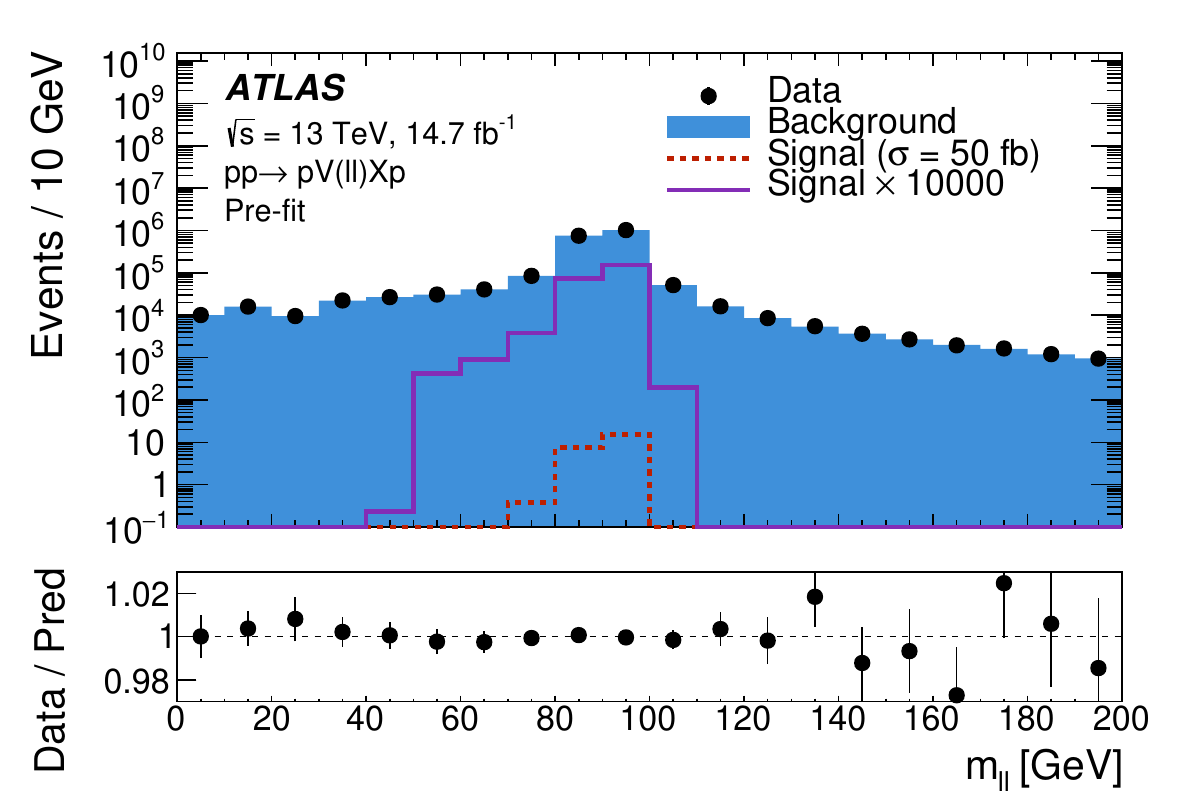}
\caption{}
\label{fig:unblinded data/bkg comp pre-sel lepton pair m comb}
\end{subfigure}
\begin{subfigure}{.45\textwidth}
\includegraphics[width=0.95\linewidth]{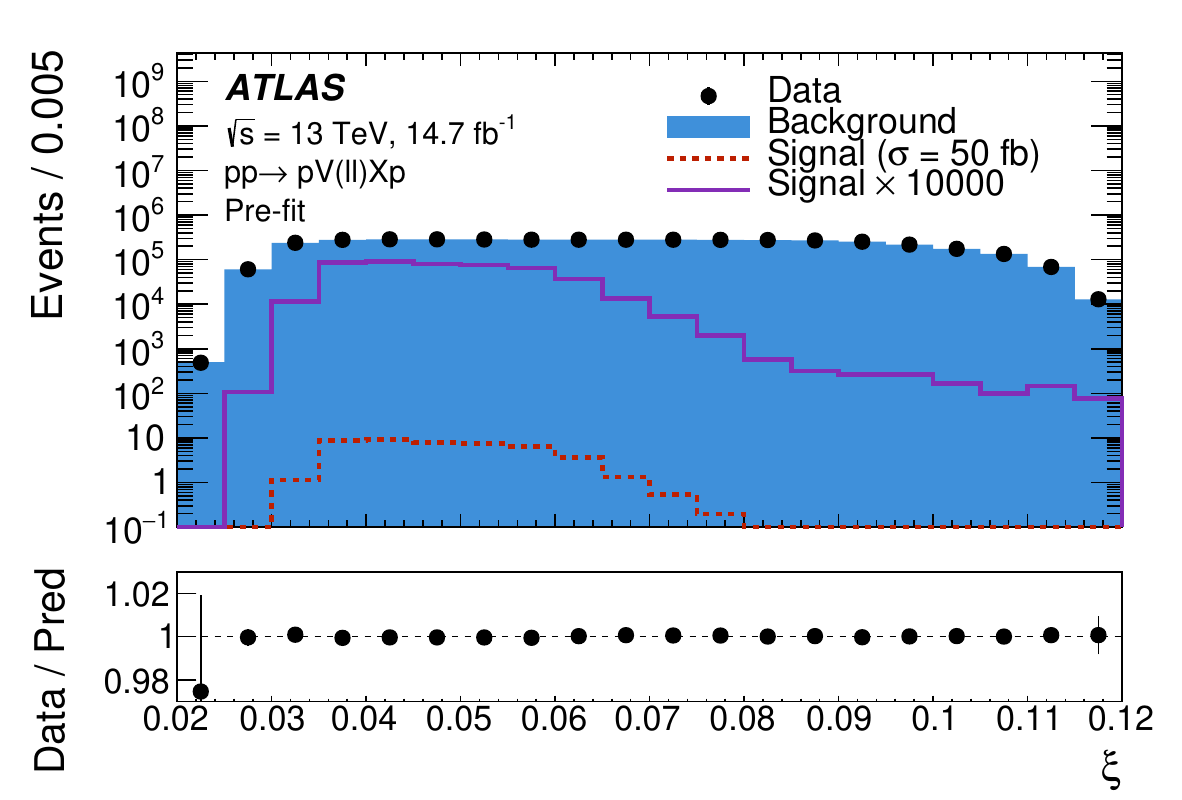}
\caption{}
\label{fig:unblinded data/bkg comp pre-sel proton xi comb}
\end{subfigure}
\begin{subfigure}{.45\textwidth}
\includegraphics[width=0.95\linewidth]{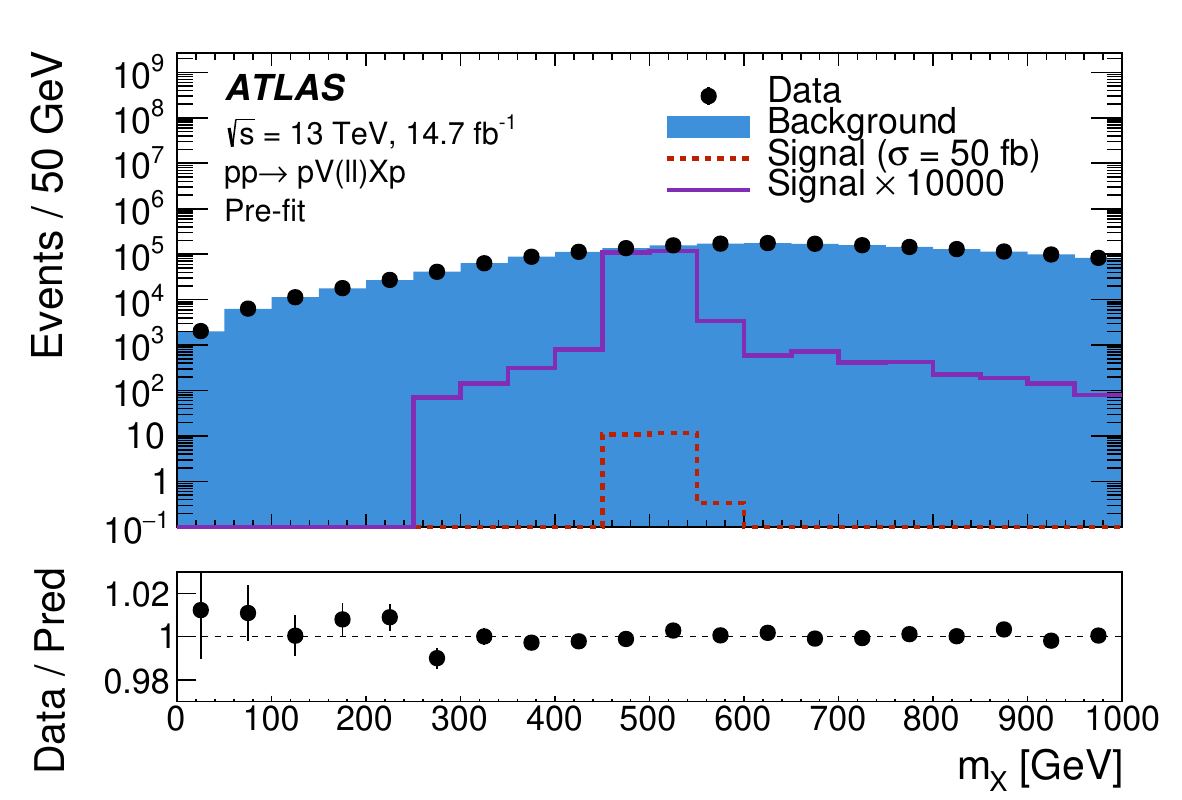}
\caption{}
\label{fig:unblinded data/bkg comp pre-sel missing mass m comb}
\end{subfigure}
\caption{Comparison between data and the data-driven background model in the combined lepton channel with only preselection requirements applied for distributions of (a) dilepton pair $p_\textrm{T}$, (b) dilepton mass $m_{\ell\ell}$, (c) proton $\xi$ and (d) missing mass $m_X$. The expectations for a signal with a hypothesised mass of $m_X = 500$~GeV from the \textsc{SuperChic} $Z+X$ model are overlaid and normalised to a cross section of 50~fb. An additional overlay of the signal scaled up by a factor of 10,000 is included to show the tails of each distribution.}
\label{fig:unblinded data/bkg comp pre-sel comb}
\end{figure}

\begin{figure}[h!]
\centering
\begin{subfigure}{.45\textwidth}
\includegraphics[width=0.95\linewidth]{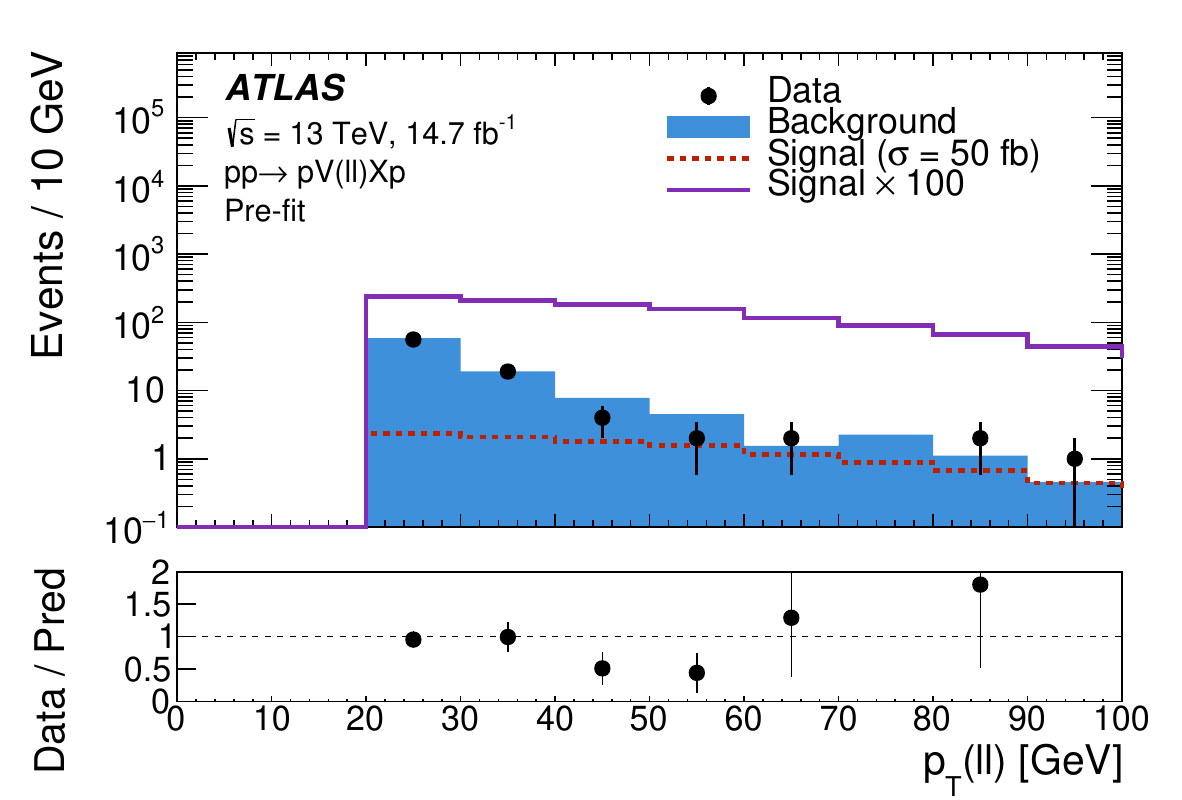}
\caption{}
\label{fig:unblinded data/bkg comp all final cuts lepton pair pt comb}
\end{subfigure}
\begin{subfigure}{.45\textwidth}
\includegraphics[width=0.95\linewidth]{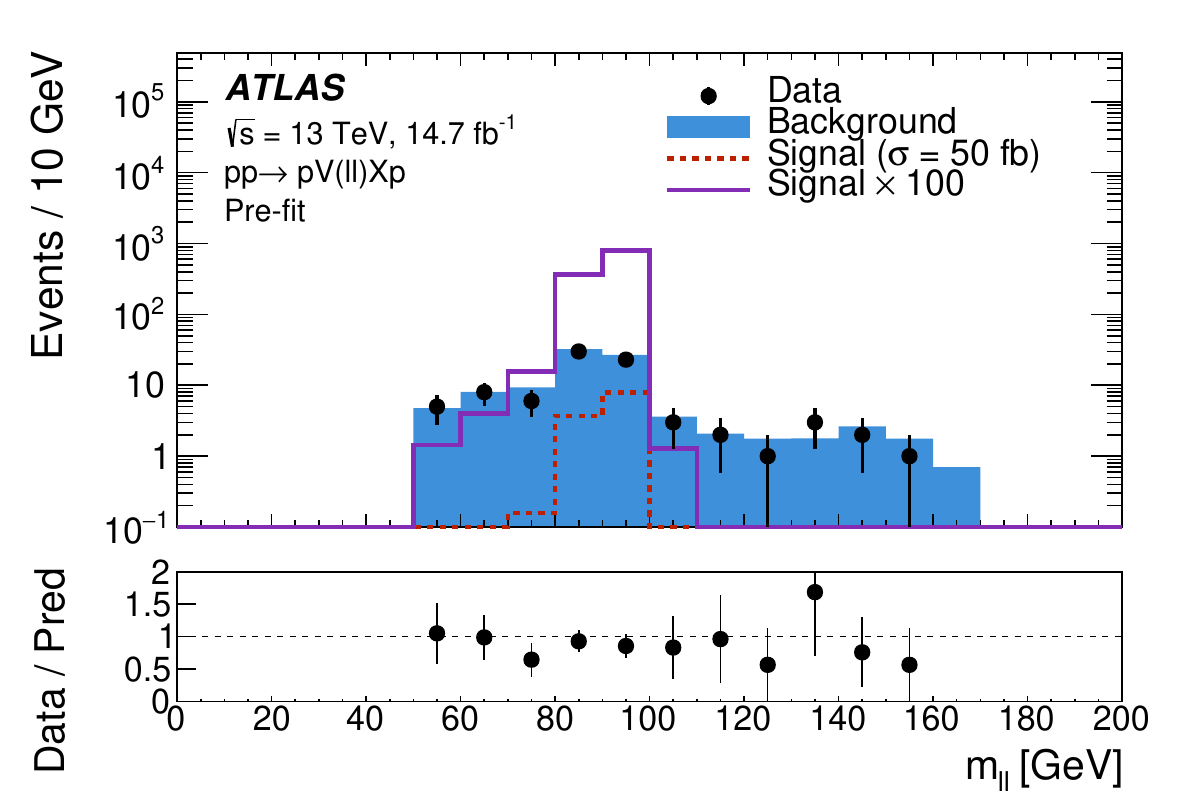}
\caption{}
\label{fig:unblinded data/bkg comp all final cuts lepton pair m comb}
\end{subfigure}
\begin{subfigure}{.45\textwidth}
\includegraphics[width=0.95\linewidth]{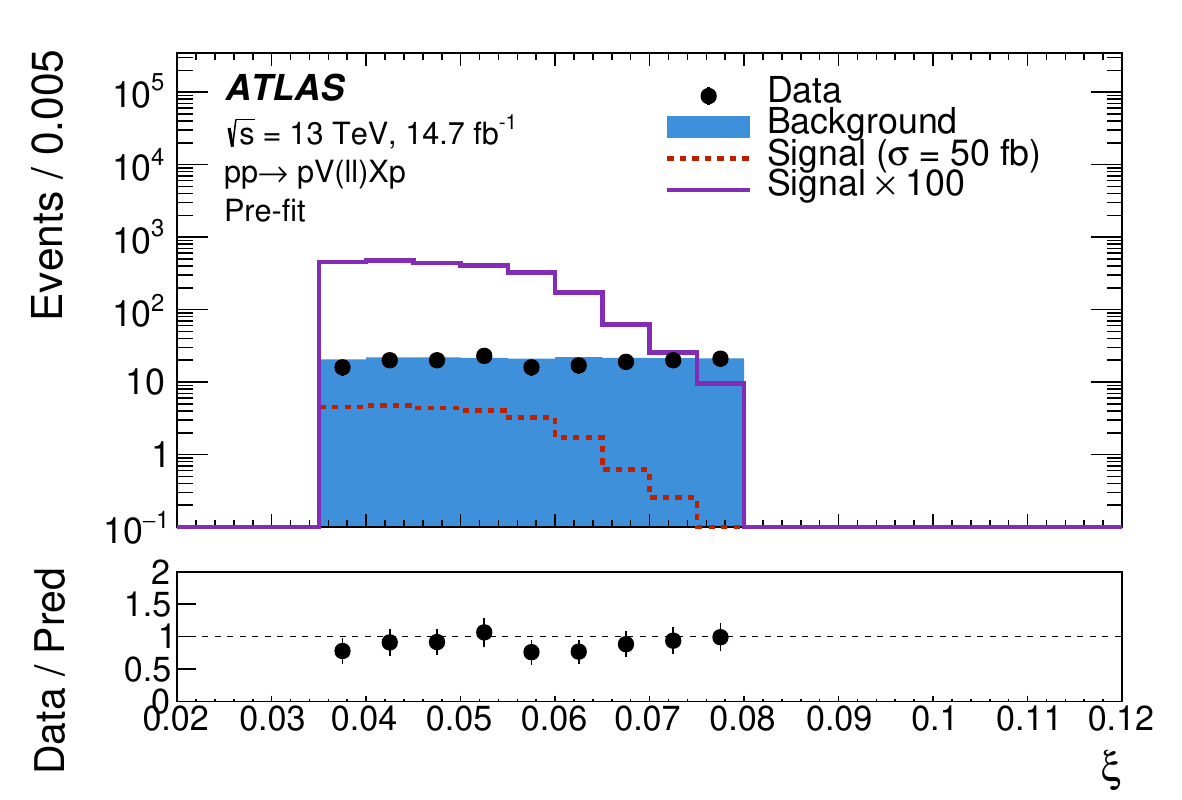}
\caption{}
\label{fig:unblinded data/bkg comp all final cuts proton xi comb}
\end{subfigure}
\begin{subfigure}{.45\textwidth}
\includegraphics[width=0.95\linewidth]{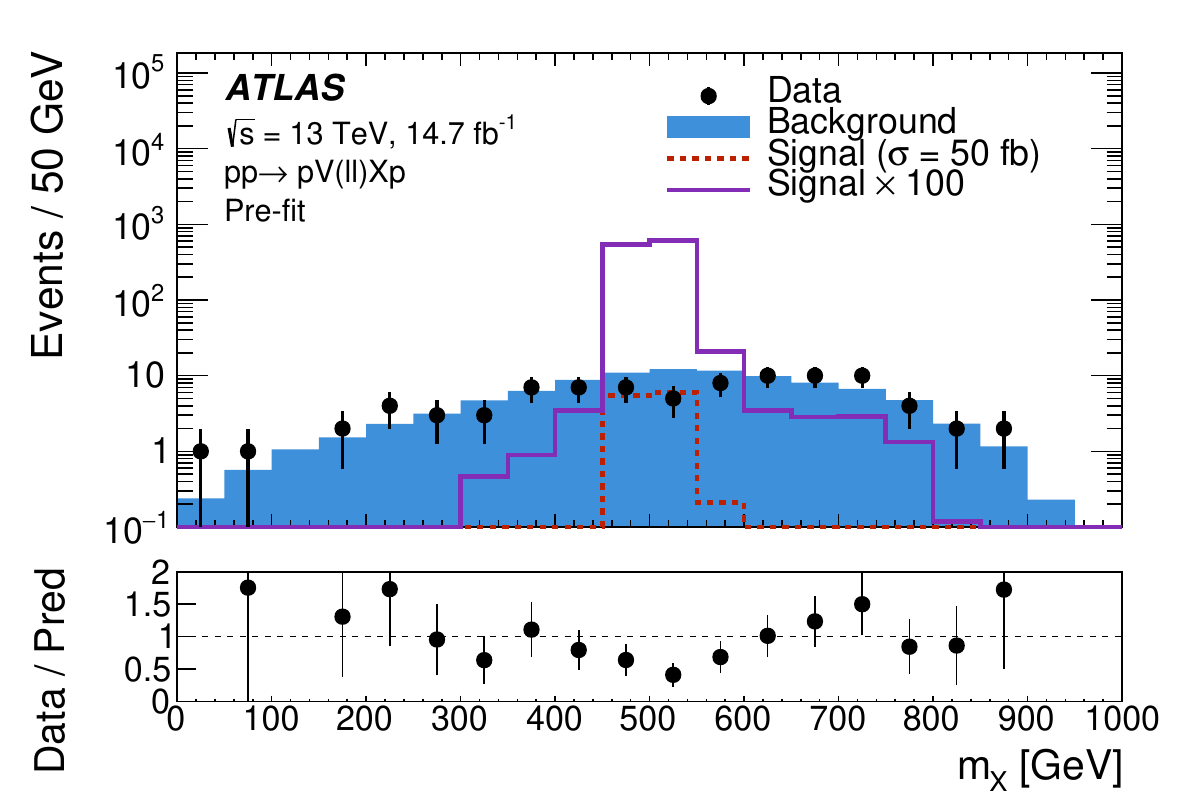}
\caption{}
\label{fig:unblinded data/bkg comp all final cuts missing mass m comb}
\end{subfigure}
\caption{Comparison between data and the data-driven background model in the combined lepton channel with the preselection and all signal region requirements applied, for distributions of (a) dilepton pair $p_\textrm{T}$, (b) dilepton mass $m_{\ell\ell}$, (c) proton $\xi$ and (d) missing mass $m_X$. The expectations for a signal with a hypothesised mass of $m_X = 500$~GeV from the \textsc{SuperChic} $Z+X$ model are overlaid and normalised to a cross-section of 50~fb. An additional overlay of the signal scaled up by a factor of 100 is included to show the tails of each distribution.}
\label{fig:unblinded data/bkg comp all final cuts comb}
\end{figure}

Figure~\ref{fig:unblinded data/bkg comp pre-sel comb} shows a comparison between the data and the data-driven background model discussed in Section~\ref{sec:background modelling} in the combined lepton channel, with only event preselection applied. Good agreement is observed between the data and the background model, and the same level of agreement is observed for the individual lepton channels. Distributions of the same observables are shown in Figure~\ref{fig:unblinded data/bkg comp all final cuts comb} for the combined lepton channel with all signal region requirements applied and no significant difference between the data and the expected background is observed.

\subsection{Model-dependent fits}
\begin{figure}[ht]
\centering
\begin{subfigure}{.49\textwidth}
\centering
\includegraphics[width=\linewidth]{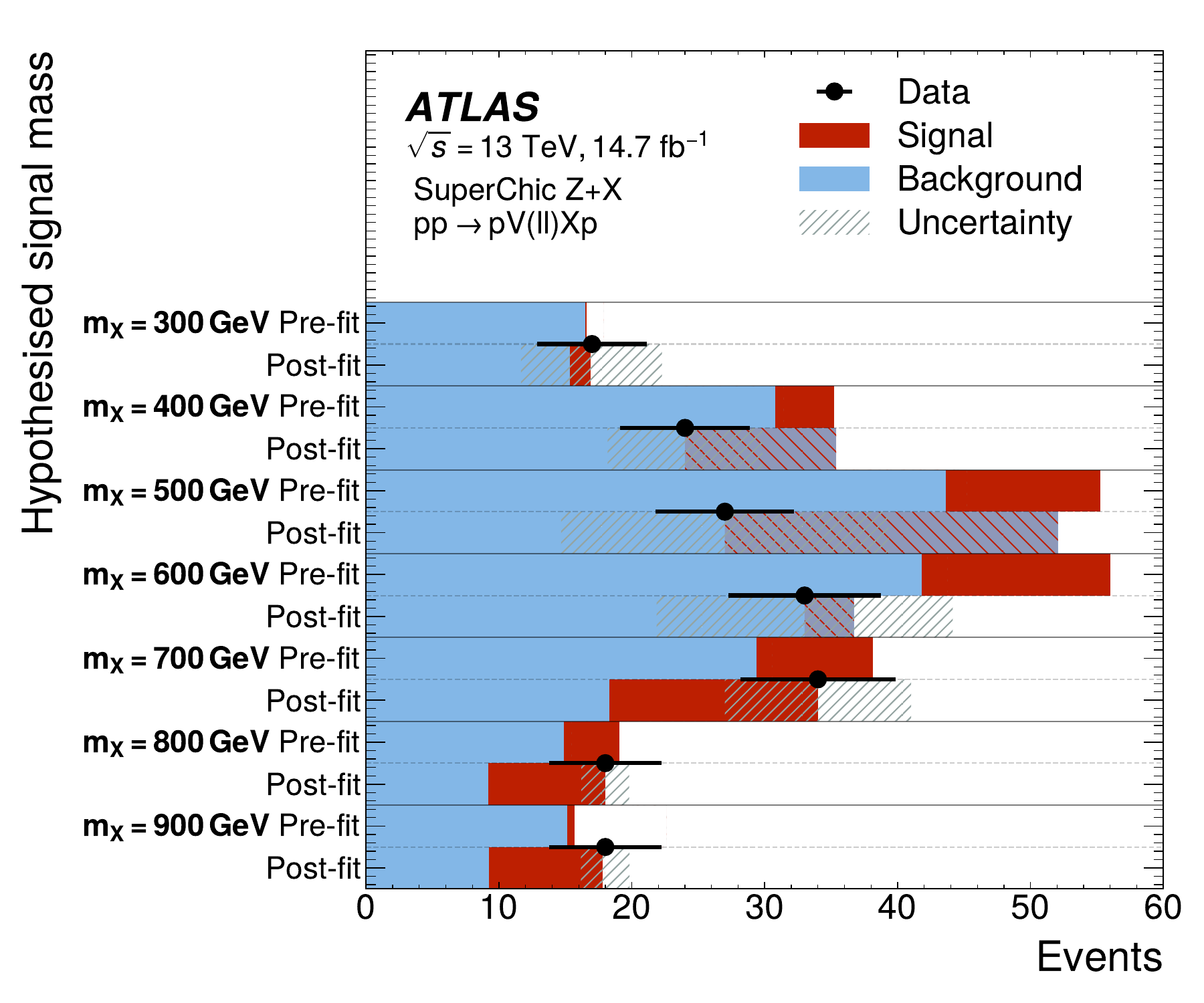}
\caption{}
\label{fig:stat sc comb fit summary}
\end{subfigure}
\begin{subfigure}{.49\textwidth}
\centering
\includegraphics[width=\linewidth]{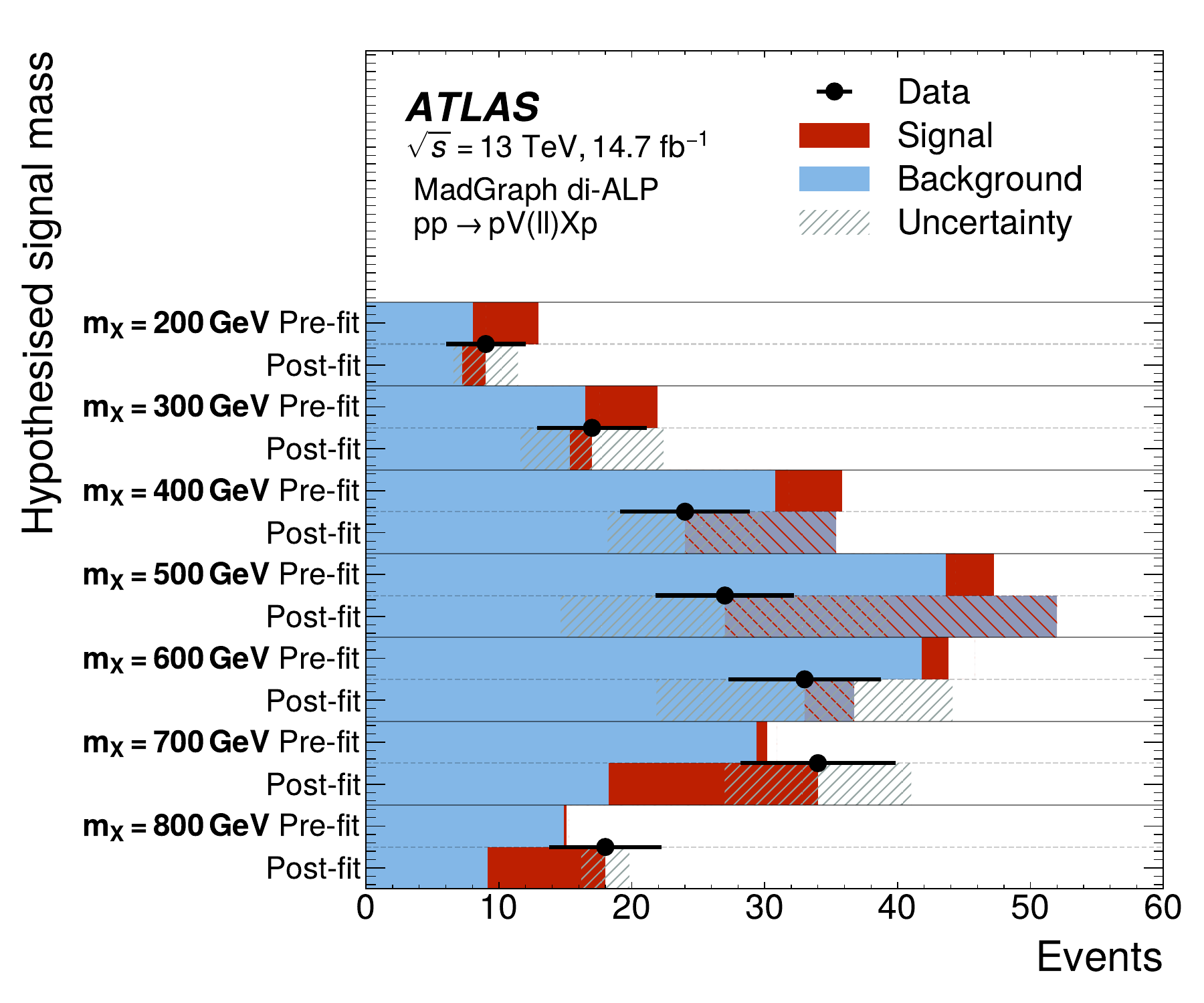}
\caption{}
\label{fig:stat mg alp comb fit summary}
\end{subfigure}
\begin{subfigure}{.49\textwidth}
\centering
\includegraphics[width=\linewidth]{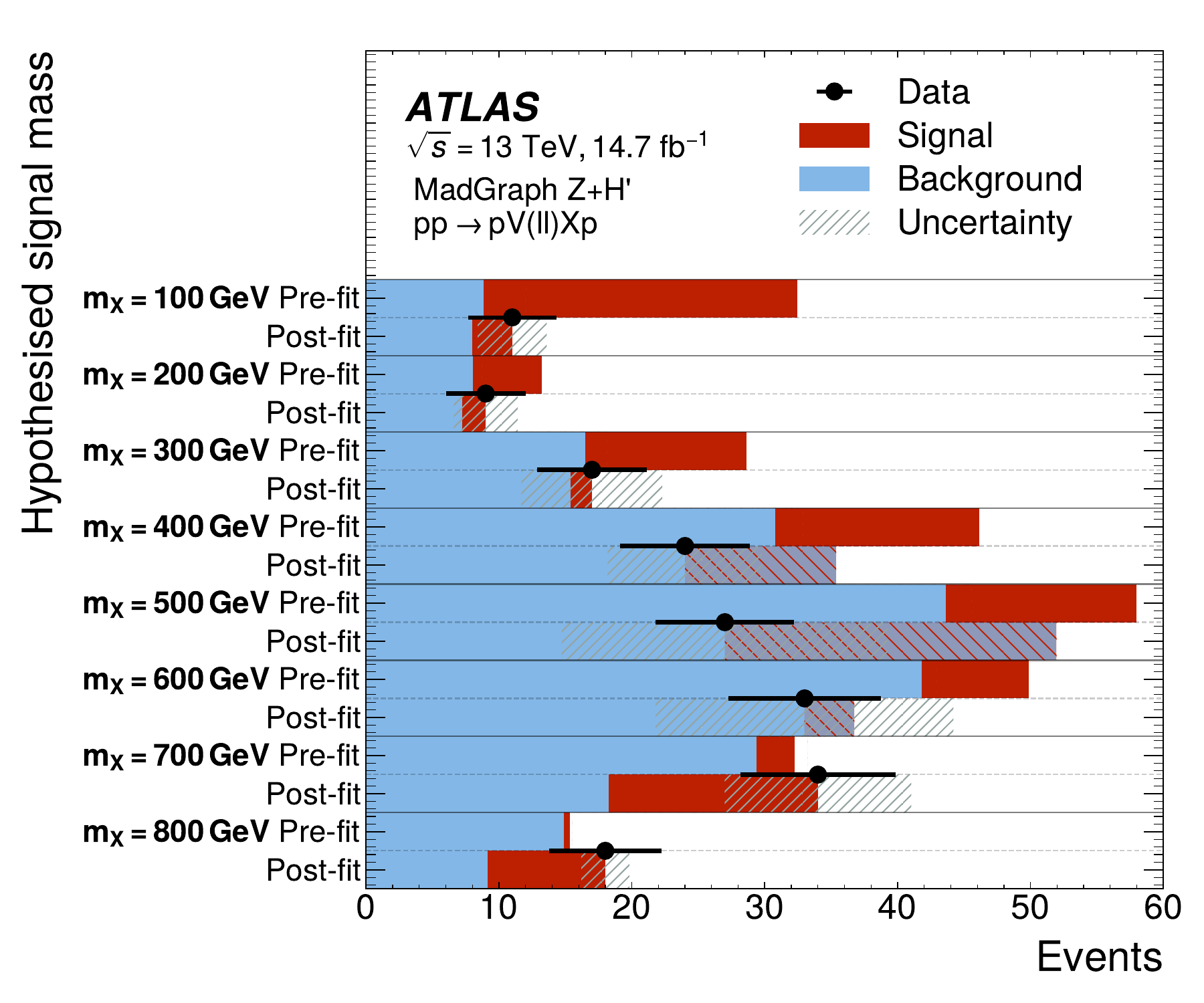}
\caption{}
\label{fig:stat mg zh comb fit summary}
\end{subfigure}
\caption{Summary of all pre- and post-fit event yields for the signal-plus-background model for each tested signal mass, in the combined lepton channel, for the (a) \textsc{SuperChic} $Z+X$, (b) \textsc{MadGraph} di-ALP and (c) \textsc{MadGraph} $Z+H'$ models. Pre-fit means before both the CR fit fixing the background normalisation, and the SR fit extracting the signal strength. Fits use a mass window of 100~GeV either side of the hypothesised signal mass, with the exception of the 100~GeV and 900~GeV models, which use larger windows of $0\leq m_X \leq 300$~GeV and $700 \leq m_X \leq 1000$~GeV respectively. All pre-fit signals are shown normalised to a cross-section of 50~fb. Negative post-fit signal strength is indicated by a red hashed area. The post-fit uncertainty corresponds to the uncertainty in the background prediction after the CR fit.}
\label{fig:stat comb fit summaries}
\end{figure}

The pre- and post-fit event yields for all tested signal masses in the combined lepton channel are shown in Figure~\ref{fig:stat comb fit summaries} for each of the three signal models, where pre-fit means before both the CR fit fixing the background normalisation, and the SR fit extracting the signal strength. The event yields correspond to the signal region, with the mass window for each signal mass set as described above.
For the pre-fit bins, all signals are normalised to a cross-section of 50~fb, to demonstrate the relative selection efficiency for each model and mass point.
After the fit, there is no observed deviation of more than $\pm2\sigma$ from the predicted event yield.

\begin{figure}[ht]
\centering
\begin{subfigure}{.325\textwidth}
\includegraphics[width=0.95\linewidth]{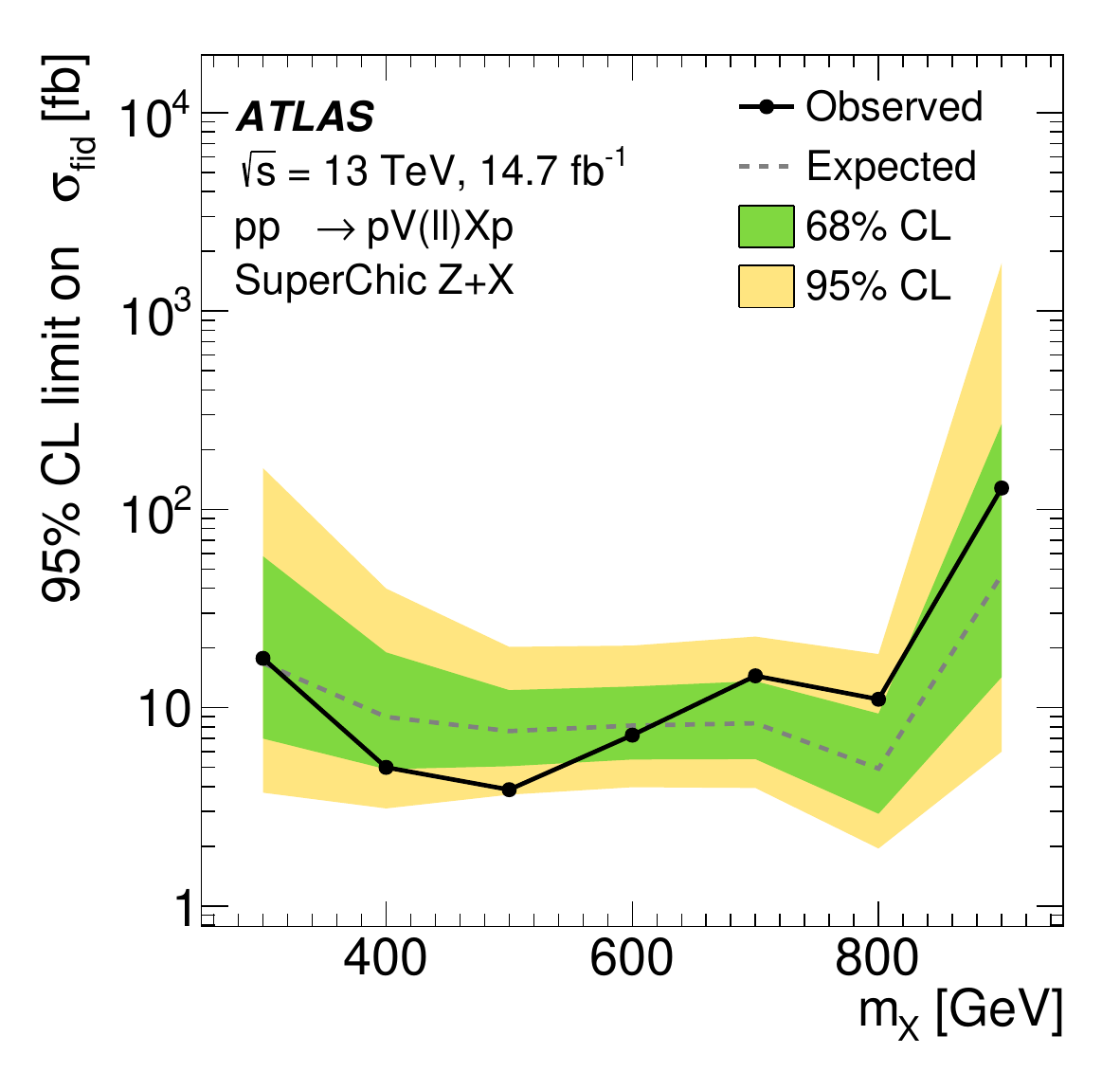}
\caption{}
\label{fig:stat sc comb limits high pt}
\end{subfigure}
\begin{subfigure}{.325\textwidth}
\includegraphics[width=0.95\linewidth]{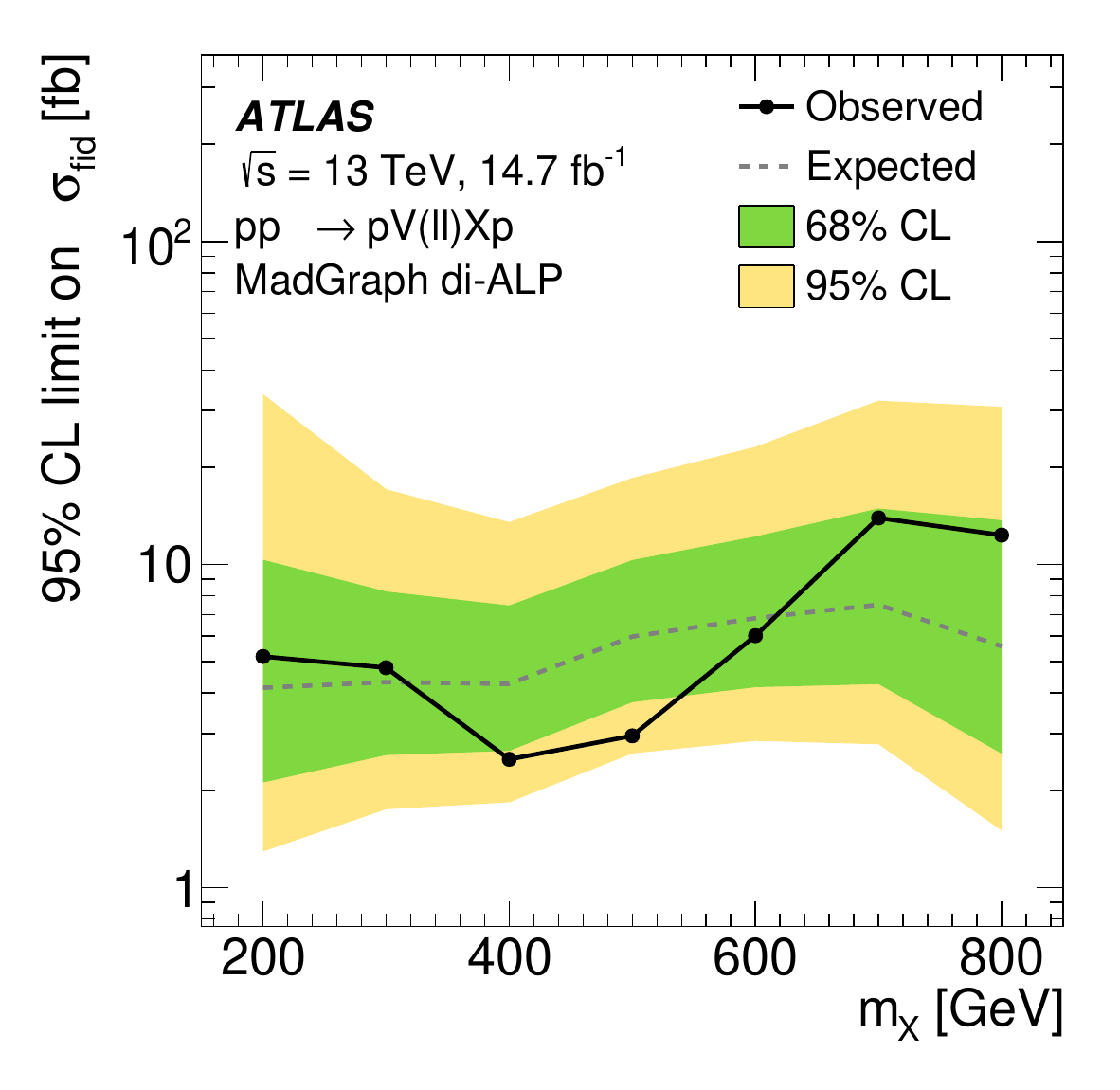}
\caption{}
\label{fig:stat mg alp comb limits high pt}
\end{subfigure}
\begin{subfigure}{.325\textwidth}
\includegraphics[width=0.95\linewidth]{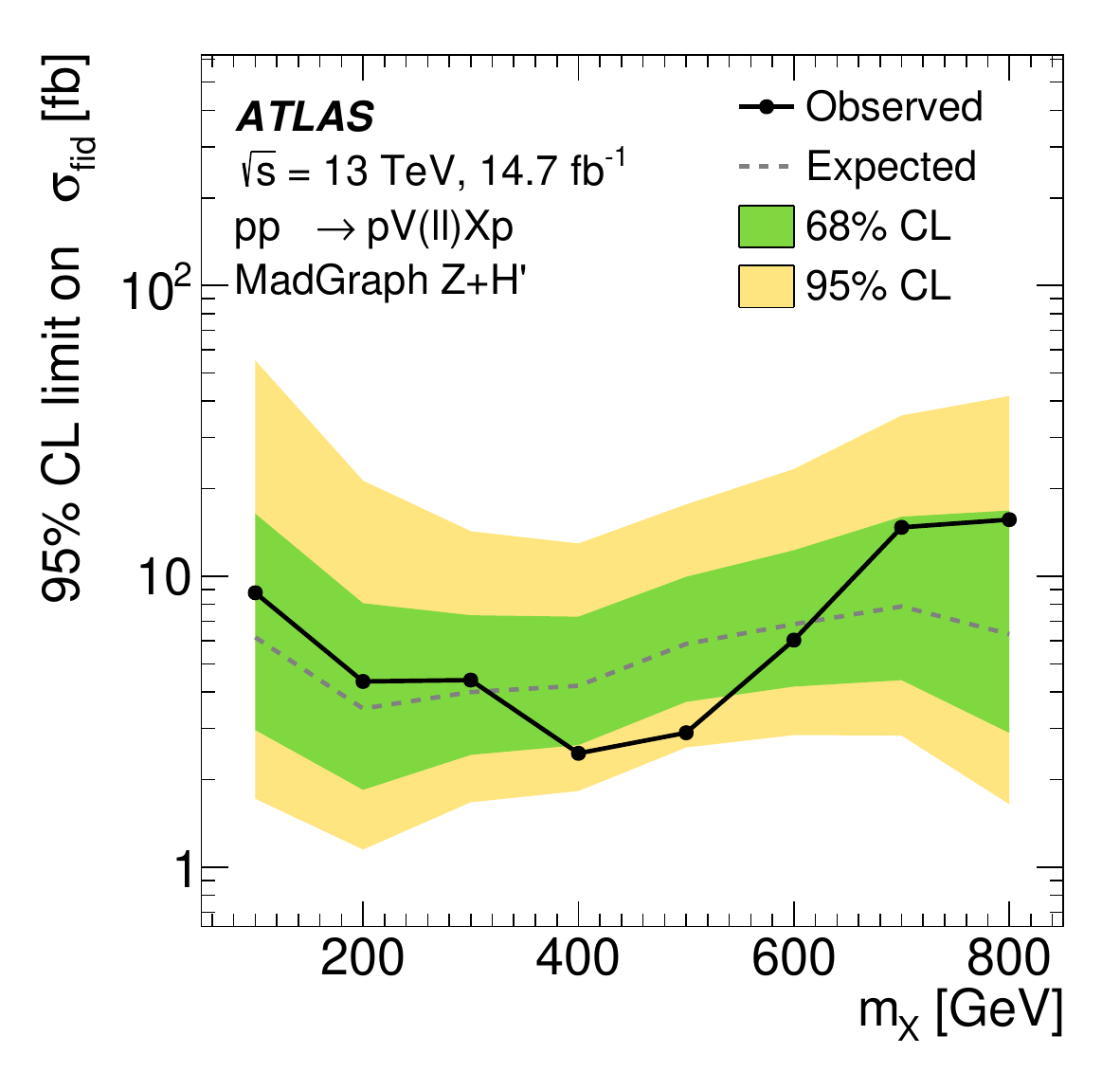}
\caption{}
\label{fig:stat mg zh comb limits high pt}
\end{subfigure}\\
\caption{Observed and expected upper limits on the fiducial cross-section set for the (a) \textsc{SuperChic} $Z+X$, (b) \textsc{MadGraph} di-ALP and (c) \textsc{MadGraph} $Z+H'$ signal models in the combined lepton channel. The $\pm1\sigma$ and $\pm2\sigma$ uncertainty bands on the excepted limit are shown in green and yellow, respectively.}
\label{fig:stat limits high pt comb}
\end{figure}

The CL$_\text{s}$ method~\cite{Read:2002hq} is used to set upper limits at 95\% confidence level (CL) on the fiducial cross-section for each signal model across the full considered mass range of up to 100--900~GeV, depending on the model. Exclusion limits are computed separately for the muon and electron channels, as well as for their statistical combination. The resulting upper limits on the signal cross-section, incorporating all systematic uncertainties and scale factors, are shown in Figure~\ref{fig:stat limits high pt comb}.
Limits of the order of 10~fb are obtained for all signal models. The dominant systematic uncertainties are found to be the uncertainties in the estimates of soft-survival probability and track-veto signal efficiency, in addition to the proton reconstruction and AFP alignment uncertainties.

The impact of systematic uncertainties on the obtained limits is evaluated by comparison with limits obtained without their inclusion. With the inclusion of systematic uncertainties the obtained upper limit increases by between 20\% and 200\% in the mass range $300\leqslant m_X \leqslant 800$~GeV common to all models. The largest impact is observed for low and high signal masses, due to the large effect of AFP-related systematic uncertainties seen in these regions.

The sensitivity of this analysis improves upon results obtained by the corresponding CMS analysis~\cite{CMS-EXO-19-009} for common mass points between 600 and 800~GeV, when adjustments are made to account for different fiducial selections and model assumptions between the two analyses. In the combined lepton channel, improvements of 770\%, 160\% and 90\% are observed for $m_X = 600$, 700 and $800$~GeV, respectively. The comparison is made using the results obtained in this analysis for the \textsc{SuperChic} $Z+X$ model, which is equivalent to the model used by CMS. The improvement comes primarily from the application of the track veto in this analysis, despite it having around 2.5 times less data than the CMS analysis. CMS was able to set limits for much higher mass values, up to 1600~GeV, due to the higher upper bound on the acceptance of their forward proton detector, the PPS, which is $\xi < 0.2$ compared to the AFP spectrometer limit of $\xi < 0.12$. The two sets of results are largely complementary, with this analysis probing much lower mass points.

\subsection{Model-independent fits}
The CL$_\text{s}$ method is also employed to set model-independent upper limits at 95\% CL on the visible cross-section $\sigma_\textrm{vis}$, defined as the cross-section times acceptance times efficiency, of BSM processes. Fits are performed on hypothesised signal masses at 50~GeV intervals in the range $50 \leqslant m_X^\textrm{sig} \leqslant 950$~GeV, with the signal yield as the parameter of interest. For these fits, the only systematic uncertainty considered is the statistical uncertainty in the background model, as no simulated models are used in these fits.

The signal region is defined for each signal mass as a 50~GeV bin centred on the signal mass. As for the model-dependent results, the control region for each mass point is defined with a minimum of 100~GeV separation from the corresponding signal region, to ensure no signal contamination is possible. Excluding this region ($m_X^\textrm{sig} - 125$~GeV to $m_X^\textrm{sig} + 125$~GeV), the control regions are defined in two slices: $0 \leqslant m_X^\textrm{CR} \leqslant 400$~GeV for $m_X^\textrm{sig} \leqslant 400$~GeV and $450 \leqslant m_X^\textrm{CR} \leqslant 1000$~GeV for $m_X^\textrm{sig} > 400$~GeV, following injection test studies. For models with $m_X^\textrm{sig} \leqslant 200$~GeV, the mass range below the signal region is excluded from the control region, due to very low statistics in this range. For the same reason the mass range above the signal region is excluded from the control regions for signal models with $m_X^\textrm{sig} \geqslant 800$~GeV. For signal masses in the range $ 250 \leqslant m_X^\textrm{sig} \leqslant 400$~GeV, the control region is extended above the 400~GeV limit to include an additional 50 GeV bin directly above the corresponding signal window for each mass point, to recover high enough statistics for the fit to converge.

\begin{figure}[ht]
\centering
\begin{subfigure}{.49\textwidth}
\includegraphics[width=0.95\linewidth]{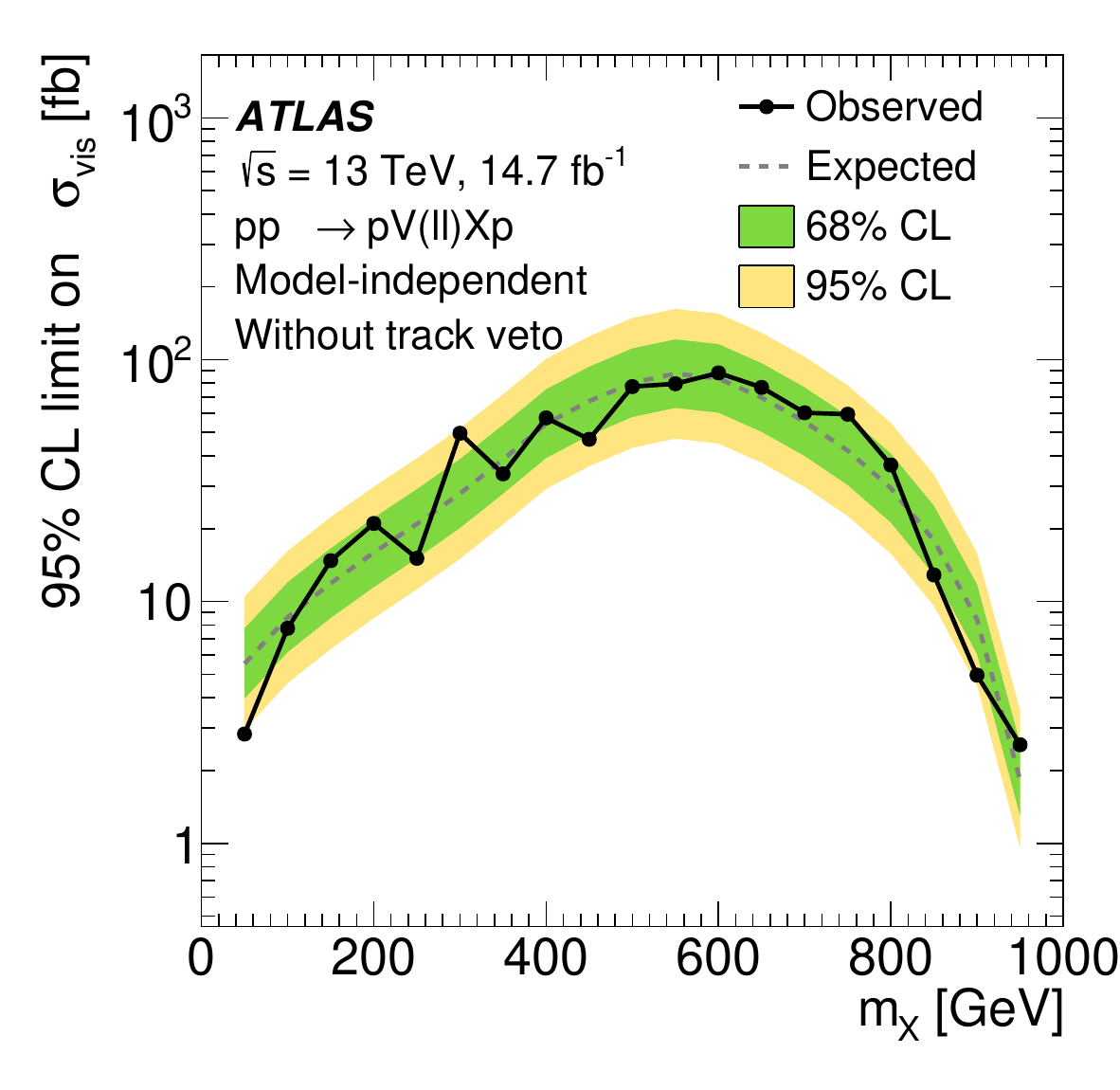}
\caption{}
\label{fig:stat model indep comb no veto}
\end{subfigure}
\begin{subfigure}{.49\textwidth}
\includegraphics[width=0.95\linewidth]{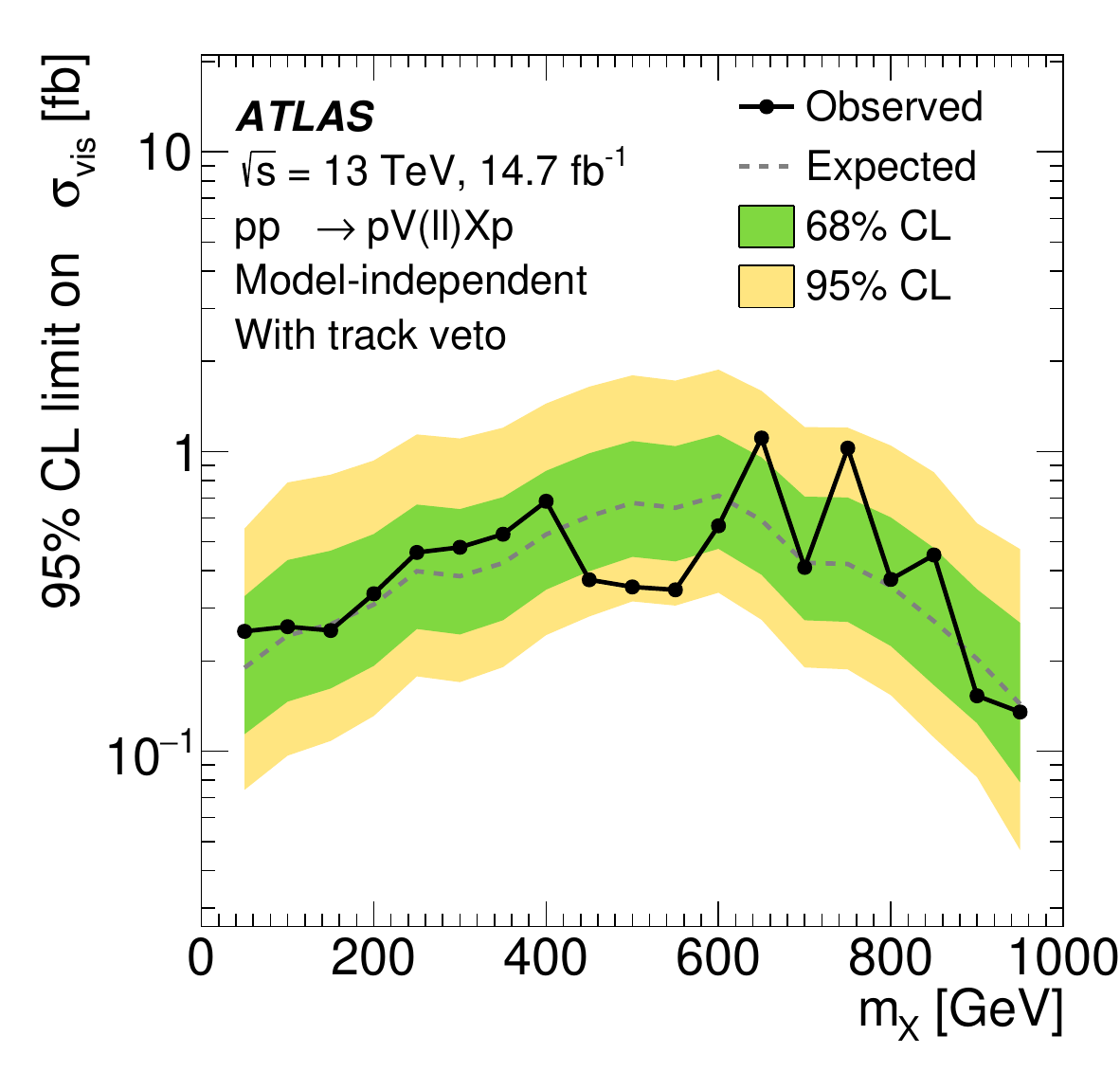}
\caption{}
\label{fig:stat model indep comb veto}
\end{subfigure}\\
\caption{Model-independent upper limits on the visible cross-section $\sigma_\textrm{vis}$ for BSM processes, set at 95\% CL using the CL$_\text{s}$ method, for the full signal selection (a) excluding the track veto and (b) including the track-veto selection. The visible cross-section is defined as the cross-section times acceptance times efficiency. Fits are performed in 50~GeV bins centred on missing masses at 50~GeV intervals in the range $50 \leqslant m_X \leqslant 950$~GeV, with the signal yield as the parameter of interest.}
\label{fig:stat model indep comb}
\end{figure}

The observed and expected 95\% CL limits $S^{0.95}_\textrm{obs}$ and $S^{0.95}_\textrm{exp}$ on the number of events from BSM processes are calculated. The $p_0$ values are also determined, which represent the probability that the SM background alone could produce at least the observed number of events purely via statistical fluctuations. The $p_0$ values are capped at 0.5 for bins where the expected yield exceeds the observed data. The corresponding significance $Z$ is given alongside each $p_0$ value. These results are presented in Tables~\ref{tab:model_ind_limits_notrk} and~\ref{tab:model_ind_limits_withtrk}, for fits performed without the track-veto selection applied, and for the full signal selection, respectively. The visible cross-section upper limits are also plotted in Figure~\ref{fig:stat model indep comb}.
\begin{table}[h]
\caption{Observed data and expected background yields for different bins of missing mass $m_X$, along with model-independent upper limits set at 95\% CL on the observed and expected number of BSM events, $S^{0.95}_\textrm{obs/exp}$, and on the effective BSM cross-section, $\sigma_\textrm{vis}$. They are obtained using a single-bin fit including the defined mass windows in the combined lepton channel for a modified signal selection without the track-veto selection applied. The $\pm 1 \sigma$ variations of $S^{0.95}_\textrm{exp}$ are provided, and the $p_0$ value of the SM-only hypothesis and its associated significance $Z$ are presented in the last column. For masses where the data yield is smaller than expected, the $p_0$ ($Z$) value is capped at $0.50$ (0.0).}
\centering
\begin{tabular}{lcc|cccc}
\toprule
Signal region [GeV] & Data & Expected & $\sigma_\textrm{vis}$ [fb] & $S^{0.95}_\textrm{obs}$ & $S^{0.95}_\textrm{exp}$ & $p_0$ ($Z$) \\
\midrule
$m_X \in [25, 75)$ & ~~1349 & $1437 \pm 16$ & ~~2.8 & ~~42 & ~~$81^{+114}_{-58}$ & 0.50 (0.0) \\
$m_X \in [75, 125)$ & ~~2994 & $3014 \pm 33$ & ~~7.7 & 114 & $126^{+177}_{-91}$ & 0.50 (0.0) \\
$m_X \in [125, 175)$ & ~~4982 & $4925 \pm 54$ & 14.8 & 217 & $175^{+244}_{-126}$ & 0.26 (0.6) \\
$m_X \in [175, 225)$ & ~~7410 & $7311 \pm 82$ & 21.0 & 309 & $234^{+327}_{-168}$ & 0.20 (0.8) \\
$m_X \in [225, 275)$ & 10499 & $10660 \pm 120$ & 15.1 & 222 & $308^{+429}_{-222}$ & 0.50 (0.0) \\
$m_X \in [275, 325)$ & 15746 & $15360 \pm 170$ & 49.6 & 730 & $411^{+571}_{-296}$ & 0.03 (1.8) \\
$m_X \in [325, 375)$ & 23142 & $23260 \pm 250$ & 33.7 & 496 & $569^{+790}_{-410}$ & 0.50 (0.0) \\
$m_X \in [375, 425)$ & 34478 & $34410 \pm 360$ & 57.5 & 845 & ~$799^{+1109}_{-576}$ & 0.50 (0.0) \\
$m_X \in [425, 475)$ & 44813 & $45390 \pm 470$ & 46.9 & 689 & ~$991^{+1376}_{-715}$ & 0.50 (0.0) \\
$m_X \in [475, 525)$ & 53807 & $53870 \pm 560$ & 77.4 & 1138~~ & $1180^{+1637}_{-851}$~ & 0.50 (0.0) \\
$m_X \in [525, 575)$ & 58189 & $58380 \pm 610$ & 79.5 & 1169~~ & $1287^{+1785}_{-929}$~ & 0.50 (0.0) \\
$m_X \in [575, 625)$ & 55107 & $55010 \pm 580$ & 88.2 & 1296~~ & $1229^{+1706}_{-887}$~ & 0.44 (0.2) \\
$m_X \in [625, 675)$ & 45876 & $45720 \pm 500$ & 76.8 & 1129~~ & $1025^{+1423}_{-739}$~ & 0.39 (0.3) \\
$m_X \in [675, 725)$ & 35959 & $35850 \pm 370$ & 60.3 & 887 & ~$815^{+1130}_{-587}$ & 0.40 (0.3) \\
$m_X \in [725, 775)$ & 26635 & $26300 \pm 270$ & 59.5 & 875 & $618^{+860}_{-445}$ & 0.15 (1.1) \\
$m_X \in [775, 825)$ & 17448 & $17300 \pm 180$ & 36.7 & 539 & $433^{+601}_{-312}$ & 0.25 (0.7) \\
$m_X \in [825, 875)$ & ~~9066 & $9200 \pm 94$ & 12.9 & 189 & $263^{+366}_{-189}$ & 0.50 (0.0) \\
$m_X \in [875, 925)$ & ~~2901 & $3006 \pm 31$ & ~~5.0 & ~~73 & $124^{+173}_{-89}$ & 0.50 (0.0) \\
$m_X \in [925, 975)$ & ~~~~178 & $165 \pm 2$ & ~~2.6 & ~~38 & ~$27^{+38}_{-19}$ & 0.16 (1.0) \\
\bottomrule
\end{tabular}
\label{tab:model_ind_limits_notrk}
\end{table}
\begin{table}[h]
\caption{Observed data and expected background yields for different bins of missing mass $m_X$, along with model-independent upper limits set at 95\% CL on the observed and expected number of BSM events, $S^{0.95}_\textrm{obs/exp}$, and on the effective BSM cross-section, $\sigma_\textrm{vis}$. They are obtained using a single-bin fit including the defined mass windows in the combined lepton channel for the full signal selection including the track-veto selection. The $\pm 1 \sigma$ variations of $S^{0.95}_\textrm{exp}$ are provided, and the $p_0$ value of the SM-only hypothesis and its associated significance $Z$ are presented in the last column. For masses where the data yield is smaller than expected, the $p_0$ ($Z$) value is capped at $0.50$ (0.0).}
\centering
\begin{tabular}{lcc|cccc}
\toprule
Signal region [GeV] & Data & Expected & $\sigma_\textrm{vis}$ [fb] & $S^{0.95}_\textrm{obs}$ & $S^{0.95}_\textrm{exp}$ & $p_0$ ($Z$) \\
\midrule
$m_X \in [25, 75)$ & 1 & $0.39 \pm 0.11$ & 0.25 & 3.7 & $3.0^{+4.8}_{-0.0}$ & 0.21 (0.8) \\
$m_X \in [75, 125)$ & 1 & $0.83 \pm 0.23$ & 0.26 & 3.8 & $3.6^{+6.4}_{-0.6}$ & 0.43 (0.2) \\
$m_X \in [125, 175)$ & 1 & $1.14 \pm 0.35$ & 0.25 & 3.7 & $3.9^{+6.9}_{-0.9}$ & 0.50 (0.0) \\
$m_X \in [175, 225)$ & 2 & $1.68 \pm 0.61$ & 0.34 & 4.9 & $4.5^{+7.8}_{-1.5}$ & 0.42 (0.2) \\
$m_X \in [225, 275)$ & 4 & $3.19 \pm 1.03$ & 0.46 & 6.8 & $5.9^{+9.8}_{-2.9}$ & 0.35 (0.4) \\
$m_X \in [275, 325)$ & 4 & $2.78 \pm 0.94$ & 0.48 & 7.0 & $5.6^{+9.5}_{-2.6}$ & 0.28 (0.6) \\
$m_X \in [325, 375)$ & 5 & $3.62 \pm 1.11$ & 0.53 & 7.8 & ~$6.2^{+10.4}_{-3.2}$ & 0.28 (0.6) \\
$m_X \in [375, 425)$ & 8 & $5.89 \pm 1.59$ & 0.68 & 10.0 & ~$7.8^{+12.7}_{-4.8}$ & 0.25 (0.7) \\
$m_X \in [425, 475)$ & 6 & $10.82 \pm 1.70$~~ & 0.37 & 5.5 & ~$8.9^{+14.5}_{-5.9}$ & 0.50 (0.0) \\
$m_X \in [475, 525)$ & 6 & $13.74 \pm 2.44$~~ & 0.35 & 5.2 & ~$9.9^{+15.9}_{-6.5}$ & 0.50 (0.0) \\
$m_X \in [525, 575)$ & 5 & $12.52 \pm 2.84$~~ & 0.35 & 5.1 & ~$9.5^{+15.3}_{-6.3}$ & 0.50 (0.0) \\
$m_X \in [575, 625)$ & 10~~ & $12.80 \pm 2.91$~~ & 0.57 & 8.3 & $10.5^{+16.8}_{-7.0}$~ & 0.50 (0.0) \\
$m_X \in [625, 675)$ & 13~~ & $6.53 \pm 1.65$ & 1.11 & 16.3 & ~$8.7^{+14.1}_{-5.7}$ & 0.04 (1.8) \\
$m_X \in [675, 725)$ & 4 & $4.21 \pm 1.01$ & 0.41 & 6.0 & ~$6.3^{+10.4}_{-3.3}$ & 0.50 (0.0) \\
$m_X \in [725, 775)$ & 10~~ & $3.42 \pm 0.70$ & 1.03 & 15.1 & ~$6.2^{+10.3}_{-3.2}$ & 0.005 (2.6)~~ \\
$m_X \in [775, 825)$ & 3 & $2.77 \pm 0.47$ & 0.37 & 5.5 & $5.2^{+8.9}_{-2.2}$ & 0.45 (0.1) \\
$m_X \in [825, 875)$ & 3 & $1.19 \pm 0.21$ & 0.45 & 6.6 & $4.0^{+7.0}_{-1.0}$ & 0.09 (1.4) \\
$m_X \in [875, 925)$ & 0 & $0.58 \pm 0.01$ & 0.20 & 3.0 & $3.0^{+5.1}_{-0.0}$ & 0.50 (0.0) \\
$m_X \in [925, 975)$ & 0 & $0.05 \pm 0.02$ & 0.20 & 3.0 & $3.0^{+4.0}_{-0.0}$ & 0.50 (0.0) \\
\bottomrule
\end{tabular}
\label{tab:model_ind_limits_withtrk}
\end{table}

The limits obtained without the track-veto selection applied are significantly less stringent, as expected. However, removing this selection eliminates the exclusivity requirement on the signal, allowing a wider range of processes to be included in a more general search.

\FloatBarrier

\section{Conclusion}
\label{sec:conclusion}
A search for central exclusive photon-induced production of a visible leptonically decaying boson $V$ along with an undetected resonance $X$ is presented. The search uses 14.7~fb$^{-1}$ of 13~TeV $pp$ collision data recorded in 2017 by the ATLAS detector at the LHC, with forward proton tagging provided by the AFP spectrometer. The $X$ system represents an undetected particle or particle system produced in association with a $Z$ boson or a short-lived axion-like particle. The missing-mass spectrum, $m_X$, is reconstructed using the measured four-momenta of the tagged protons and central lepton pair, allowing a generic search enabled by the data collected by the AFP spectrometer.

No significant excess over the Standard Model expectation is observed. The results are interpreted as upper limits on the fiducial cross-section for three signal models: two simplified benchmarks involving a $Z$ boson ($Z+X$ and $Z+H'$), and a simplified BSM scenario involving axion-like particles (di-ALP), with observed fiducial cross-section limits going down to 3.9~fb, 2.5~fb and 2.5~fb, respectively. The results obtained  for the \textsc{SuperChic} $Z+X$ model improve on those from a previous study performed by CMS, with an equivalent signal model, for several common mass points between 600 and 800~GeV. This improvement is achieved mainly by using a track veto, which strongly suppresses backgrounds by exploiting central-detector activity.

This is the first ATLAS analysis to use AFP data in conjunction with the missing-mass method, and the first to apply a central-track veto in this context. The use of multiple signal models extends the reach of the search across a broad kinematic phase space. Additional model-independent visible cross-section limits of the order of 1~fb, obtained by applying the full signal selection, ensure a more general applicability of this result in the search of BSM physics.

\section*{Acknowledgements}
%

%

%
%

%
%

We thank CERN for the very successful operation of the LHC and its injectors, as well as the support staff at
CERN and at our institutions worldwide without whom ATLAS could not be operated efficiently.

The crucial computing support from all WLCG partners is acknowledged gratefully, in particular from CERN, the ATLAS Tier-1 facilities at TRIUMF/SFU (Canada), NDGF (Denmark, Norway, Sweden), CC-IN2P3 (France), KIT/GridKA (Germany), INFN-CNAF (Italy), NL-T1 (Netherlands), PIC (Spain), RAL (UK) and BNL (USA), the Tier-2 facilities worldwide and large non-WLCG resource providers. Major contributors of computing resources are listed in Ref.~\cite{ATL-SOFT-PUB-2026-001}.

We gratefully acknowledge the support of ANPCyT, Argentina; YerPhI, Armenia; ARC, Australia; BMWFW and FWF, Austria; ANAS, Azerbaijan; CNPq and FAPESP, Brazil; NSERC, NRC and CFI, Canada; CERN; ANID, Chile; CAS, MOST and NSFC, China; Minciencias, Colombia; MEYS CR, Czech Republic; DNRF and DNSRC, Denmark; IN2P3-CNRS and CEA-DRF/IRFU, France; SRNSFG, Georgia; BMFTR, HGF and MPG, Germany; GSRI, Greece; RGC and Hong Kong SAR, China; ICHEP and Academy of Sciences and Humanities, Israel; INFN, Italy; MEXT and JSPS, Japan; CNRST, Morocco; NWO, Netherlands; RCN, Norway; MNiSW, Poland; FCT, Portugal; MNE/IFA, Romania; MSTDI, Serbia; MSSR, Slovakia; ARIS and MVZI, Slovenia; DSI/NRF, South Africa; MICIU/AEI, Spain; SRC and Wallenberg Foundation, Sweden; SERI, SNSF and Cantons of Bern and Geneva, Switzerland; NSTC, Taipei; TENMAK, T\"urkiye; STFC/UKRI, United Kingdom; DOE and NSF, United States of America.

Individual groups and members have received support from BCKDF, CANARIE, CRC and DRAC, Canada; CERN-CZ, FORTE and PRIMUS, Czech Republic; COST, ERC, ERDF, Horizon 2020 and Marie Sk{\l}odowska-Curie Actions, European Union; Investissements d'Avenir Labex, Investissements d'Avenir Idex and ANR, France; DFG and AvH Foundation, Germany; Herakleitos, Thales and Aristeia programmes co-financed by EU-ESF and the Greek NSRF, Greece; BSF-NSF and MINERVA, Israel; NCN and NAWA, Poland; La Caixa Banking Foundation, CERCA and AGAUR programs from Generalitat de Catalunya and PROMETEO and GenT Programmes Generalitat Valenciana, Spain; G\"{o}ran Gustafssons Stiftelse, Sweden; The Royal Society and Leverhulme Trust, United Kingdom; Eric and Wendy Schmidt Fund for Strategic Innovation, United States of America.

In addition, individual members wish to acknowledge support from Chile: Agencia Nacional de Investigaci\'on y Desarrollo (ANID FONDECYT reg. 1230987, FONDECYT 1230812, FONDECYT 1240864, Fondecyt 3240661, Fondecyt Regular 1240721); China: Chinese Ministry of Science and Technology (MOST-2023YFA1605700, MOST-2023YFA1609300), National Natural Science Foundation of China (NSFC - 12175119, NSFC 12275265); Czech Republic: Czech Science Foundation (GACR - 24-11373S), Ministry of Education Youth and Sports (ERC-CZ-LL2327, FORTE CZ.02.01.01/00/22\_008/0004632), PRIMUS Research Programme (PRIMUS/21/SCI/017); EU: H2020 European Research Council (ERC - 101002463); European Union: European Research Council (BARD No. 101116429, ERC - 948254, ERC 101089007), European Regional Development Fund (HE COFUND GA No.101081355, ERDF), European Union, Future Artificial Intelligence Research (FAIR-NextGenerationEU PE00000013), Marie Sklodowska-Curie Actions (GAP-101168829); France: Agence Nationale de la Recherche (ANR-21-CE31-0013, ANR-21-CE31-0022, ANR-22-EDIR-0002, ANR-24-CE31-0504-01); Germany: Deutsche Forschungsgemeinschaft (DFG - 469666862, DFG - CR 312/5-2); China: Research Grants Council (GRF); Italy: Ministero dell'Università e della Ricerca (NextGenEU 153D23001490006 M4C2.1.1, NextGenEU I53D23000820006 M4C2.1.1, NextGenEU I53D23001490006 M4C2.1.1, SOE2024\_0000023); Japan: Japan Society for the Promotion of Science (JSPS KAKENHI  JP25H0063, JSPS KAKENHI JP22H01227, JSPS KAKENHI JP22H04944, JSPS KAKENHI JP22KK0227, JSPS KAKENHI JP24K23939, JSPS KAKENHI JP24KK0251, JSPS KAKENHI JP25H00650, JSPS KAKENHI JP25H01291, JSPS KAKENHI JP25K01011, JSPS KAKENHI JP25K01023); Norway: Research Council of Norway (RCN-314472); Poland: Ministry of Science and Higher Education (IDUB AGH, POB8, D4 no 9722), Polish National Science Centre (NCN 2021/42/E/ST2/00350, NCN OPUS 2023/51/B/ST2/02507, NCN OPUS nr 2022/47/B/ST2/03059, NCN UMO-2019/34/E/ST2/00393, UMO-2022/47/O/ST2/00148, UMO-2023/49/B/ST2/04085, UMO-2023/51/B/ST2/00920, UMO-2024/53/N/ST2/00869); Portugal: Foundation for Science and Technology (FCT); Spain: Agència de Gestió d'Ajuts Universitaris i de Recerca. (AGAUR - 2023 BP 00141), Generalitat Valenciana (ASFAE/2022/008), Ministry of Science and Innovation (RYC2019-028510-I, RYC2020-030254-I, RYC2021-031273-I, RYC2022-038164-I), Ministerio de Ciencia, Innovación y Universidades/Agencia Estatal de Investigaci\'on (EU NextGenerationEU (PRTR-C17.I1), PID2022-142604OB-C22); Sweden: Carl Trygger Foundation (Carl Trygger Foundation CTS 22:2312), Swedish Research Council (Swedish Research Council 2023-04654, VR 2021-03651, VR 2022-03845, VR 2022-04683, VR 2023-03403, VR 2024-05451), Knut and Alice Wallenberg Foundation (KAW 2018.0458, KAW 2022.0358, KAW 2023.0366); Switzerland: Swiss National Science Foundation (SNSF - PCEFP2\_194658); United Kingdom: The Binks Trust, Royal Society (NIF-R1-231091); United States of America: U.S. Department of Energy (ECA DE-AC02-76SF00515), John Templeton Foundation (John Templeton Foundation 63206), Neubauer Family Foundation.

%
%


%
%
%
%
%
%
%
%
%

%
%
%
%
%
%

%
%

%
%
%
\printbibliography
\clearpage
 
\begin{flushleft}
\hypersetup{urlcolor=black}
{\Large The ATLAS Collaboration}

\bigskip

\AtlasOrcid[0000-0002-6665-4934]{G.~Aad}$^\textrm{\scriptsize 103}$,
\AtlasOrcid[0000-0001-7616-1554]{E.~Aakvaag}$^\textrm{\scriptsize 17}$,
\AtlasOrcid[0000-0002-5888-2734]{B.~Abbott}$^\textrm{\scriptsize 121}$,
\AtlasOrcid[0000-0002-0287-5869]{S.~Abdelhameed}$^\textrm{\scriptsize 84b}$,
\AtlasOrcid[0000-0002-1002-1652]{K.~Abeling}$^\textrm{\scriptsize 55}$,
\AtlasOrcid[0000-0001-5763-2760]{N.J.~Abicht}$^\textrm{\scriptsize 49}$,
\AtlasOrcid[0000-0002-8496-9294]{S.H.~Abidi}$^\textrm{\scriptsize 30}$,
\AtlasOrcid[0009-0003-6578-220X]{M.~Aboelela}$^\textrm{\scriptsize 45}$,
\AtlasOrcid[0000-0002-9987-2292]{A.~Aboulhorma}$^\textrm{\scriptsize 36e}$,
\AtlasOrcid[0000-0001-5329-6640]{H.~Abramowicz}$^\textrm{\scriptsize 155}$
% ...
\bigskip
\\

% ...
$^{17}$Department for Physics and Technology, University of Bergen, Bergen; Norway.\\
% ...
$^{30}$Physics Department, Brookhaven National Laboratory, Upton NY; United States of America.\\
% ...
$^{36}$$^{(a)}$Facult\'e des Sciences Ain Chock, Universit\'e Hassan II de Casablanca;$^{(b)}$Facult\'{e} des Sciences, Universit\'{e} Ibn-Tofail, K\'{e}nitra;$^{(c)}$Facult\'e des Sciences Semlalia, Universit\'e Cadi Ayyad, LPHEA-Marrakech;$^{(d)}$LPMR, Facult\'e des Sciences, Universit\'e Mohamed Premier, Oujda;$^{(e)}$Facult\'e des sciences, Universit\'e Mohammed V, Rabat;$^{(f)}$Institute of Applied Physics, Mohammed VI Polytechnic University, Ben Guerir; Morocco.\\
% ...
$^{45}$Physics Department, Southern Methodist University, Dallas TX; United States of America.\\
% ...
$^{49}$Fakult\"{a}t Physik, Technische Universit{\"a}t Dortmund, Dortmund; Germany.\\
% ...
$^{55}$II. Physikalisches Institut, Georg-August-Universit\"{a}t G\"ottingen, G\"ottingen; Germany.\\
% ...
$^{84}$$^{(a)}$Khalifa University of Science and Technology, Abu Dhabi;$^{(b)}$New York University Abu Dhabi, Abu Dhabi;$^{(c)}$United Arab Emirates University, Al Ain;$^{(d)}$University of Sharjah, Sharjah; United Arab Emirates.\\
% ...
$^{103}$CPPM, Aix-Marseille Universit\'e, CNRS/IN2P3, Marseille; France.\\
% ...
$^{121}$Homer L. Dodge Department of Physics and Astronomy, University of Oklahoma, Norman OK; United States of America.\\
% ...
$^{155}$Raymond and Beverly Sackler School of Physics and Astronomy, Tel Aviv University, Tel Aviv; Israel.\\
% ...
$^{a}$ Also at Affiliated with an institute formerly covered by a cooperation agreement with CERN.\\
$^{b}$ Also at An-Najah National University, Nablus; Palestine.\\
$^{c}$ Also at Borough of Manhattan Community College, City University of New York, New York NY; United States of America.\\
$^{d}$ Also at Center for Interdisciplinary Research and Innovation (CIRI-AUTH), Thessaloniki; Greece.\\
$^{e}$ Also at Centre of Physics of the Universities of Minho and Porto (CF-UM-UP); Portugal.\\
$^{f}$ Also at CERN, Geneva; Switzerland.\\
% ...

\end{flushleft}


%

%
%

%

%
%
%
%

\end{document}